\documentclass[a4paper]{article}
\usepackage[top=2.5cm,left=2.5cm,right=2.5cm,bottom=1.5cm]{geometry}
\usepackage{authblk}             % affiliations for authors
\usepackage{cite}
\usepackage[obeyspaces]{url}     % [show spaces in url] to replace verb
\usepackage[parfill]{parskip}    % empty line between paragraphs, no indentation
\usepackage{color}
\usepackage{graphicx}
\usepackage{amsmath}
\usepackage{amsfonts}
\usepackage[utf8]{inputenc}      % for slovak end german characters
\usepackage{mathtools}  
\usepackage{calrsfs}             % for nice H and L and ugly M
\usepackage{xfrac}               % nice fractions
\usepackage{braket}              % bra-ket notation

\usepackage{wrapfig}
\usepackage{commath}
\usepackage{bm}
\usepackage{units}
\usepackage[usenames,dvipsnames,svgnames,table]{xcolor}

\DeclareMathOperator{\sech}{sech}
\DeclareMathOperator{\gd}{gd}

\newcommand{\red}[1]{\textcolor{black}{#1}}

\newcommand{\pa}{\partial_1}
\newcommand{\pb}{\partial_2}

\newcommand{\pt}{\partial_t}

\newcommand{\prv}{\nabla_{\vec{r}}}

\newcommand{\vp}{\varphi}

\newcommand{\matrixel}[3]{\left< #1 \vphantom{#2#3} \right| #2 \left| #3 \vphantom{#1#2} \right>}

\renewcommand{\vec}{\bm}

\title{On the Geometric Potential and the Relationship between the Exact Electron Factorization and Density Functional Theory}
\author[1]{Jakub Koc\'ak}
\author[2]{Eli Kraisler}
\author[1]{Axel Schild}
\affil[1]{Laboratorium für Physikalische Chemie, ETH Z\"urich, Vladimir-Prelog-Weg 2, 8093 Z\"urich, Switzerland}
\affil[2]{Fritz Haber Center for Molecular Dynamics, Institute of Chemistry, The Hebrew University of Jerusalem, 91904 Jerusalem, Israel}
\date{}

\begin{document}
  
  \maketitle
  
  \begin{abstract}
    There are different ways to obtain an exact one-electron theory for a many-electron system, and the exact electron factorization (EEF) is one of them.
    In the EEF, the Schr\"odinger equation for one electron in the environment of other electrons is constructed.
    The environment provides the potentials that appear in this equation:
    A scalar potential $v^{\rm H}$ representing the energy of the environment and another scalar potential $v^{\rm G}$ as well as a vector potential that have geometric meaning.
    By replacing the interacting many-electron system with the non-interacting Kohn-Sham (KS) system, we show how the EEF is related to density functional theory (DFT) and we interpret the Hartree-exchange-correlation potential as well as the Pauli potential in terms of the EEF.
    In particular, we show that from the EEF viewpoint, the Pauli potential does not represent the difference between a fermionic and a bosonic non-interacting system, but that it corresponds to $v^{\rm G}$ and partly to $v^{\rm H}$ for the (fermionic) KS system.
    We then study the meaning of $v^{\rm G}$ in detail:
    Its geometric origin as a metric measuring the change of the environment is presented.
    Additionally, its behavior for a simple model of a homo- and heteronucler diatomic is investigated and interpreted with the help of a two-state model.
    In this way, we provide a physical interpretation for the one-electron potentials that appear in the EEF and in DFT.
  \end{abstract}
%   \begin{center}
%     Corresponding author: Axel Schild (axel.schild@phys.chem.ethz.ch)
%   \end{center}
  
  \section{Introduction}
  \label{sec:intro}
  
  The quantum-mechanical solution of the many-electron problem is difficult but necessary to determine the properties of molecules and materials as well as to predict the outcomes of chemical reactions \cite{dirac1929,motta2017}.
  Density functional theory (DFT) \cite{hohenberg1964} is a highly successful approach to solve this problem \cite{dreizler1990,ullrich2013}.
  The central idea of the most widely used variant of DFT, KS-DFT \cite{kohn1965}, is to map an interacting many-electron system to a fictitious system of non-interacting electrons, the Kohn-Sham (KS) system, such that both systems have the same one-electron density $\rho(\vec{r})$ for an electron with coordinates $\vec{r}$.
  As the electrons in the KS system are non-interacting, the many-electron problem is effectively reduced to a one-electron problem.
  To determine the KS system, the one-electron KS potential $v^{\rm KS}(\vec{r})$ is needed, which is typically treated as a functional of $\rho$ or of the KS orbitals, i.e., of the eigenfunctions of $v^{\rm KS}$.
  The functional dependence is not completely known, but suitable approximations allow to answer many questions of physical and chemical relevance \cite{mohr2015}.
  
  However, the amplitude $\sqrt{\rho(\vec{r})}$ of the one-electron density is itself an eigenstate of a one-electron Schr\"odinger equation with an effective potential $v(\vec{r})$ \cite{levy1984,march1986,march1987,levy1988}.
  This fact is the basis of orbital-free DFT (OF-DFT), a method that may be computationally very efficient for large systems if suitable approximations are found \cite{wang2002,karasiev2012}.
  In the theory of OF-DFT, the KS system is usually considered as a reference and the effective potential is written as 
  \begin{align}
    v(\vec{r}) = v^{\rm P}(\vec{r}) + v^{\rm KS}(\vec{r}), \label{eq:pauli}
  \end{align}
  where the Pauli potential $v^{\rm P}$ and its properties have recently attracted some attention\cite{deb1983,perdew2007,karasiev2009,karasiev2013,karasiev2015,finzel2016,finzel2017,constantin2018,finzel2018a,finzel2018b,finzel2019,smiga2020,kraisler2020}.
  
  Although the usual approach in OF-DFT is to view the effective potential $v$ as a functional of the one-electron density, the formalism has an interesting advantage:
  In contrast to the KS potential $v^{\rm KS}$, a general equation for the potential $v$ in terms of quantities derived from the many-electron wavefunction can be given explicitly \cite{hunter1986}.
  To obtain this equation, the $N$-electron wavefunction is written as a product of a marginal wavefunction ($\sqrt{\rho(\vec{r})}$ up to a possibly $\vec{r}$-dependent phase) and a conditional wavefunction (defined below).
  The marginal wavefunction depends only on the coordinate of one electron and is an eigenstate of the same one-electron Schr\"odinger equation that is the basic equation of OF-DFT \cite{kraisler2020rev}.
  The conditional wavefunction depends on the coordinates of $N-1$ electrons, and, also, parametrically on the coordinates of the remaining electron of the $N$-electron system.
  The potential $v$ appearing in the one-electron Schr\"odinger equation of OF-DFT is a functional of the conditional wavefunction.
  
  The separation of a wavefunction into a marginal and a conditional part was first considered for the electron-nuclear problem \cite{hunter1975} and has subsequently been transferred to the many-electron problem \cite{hunter1986}, which lead to first studies of the properties of and the connections between $v$, $v^{\rm KS}$, and $v^{\rm P}$, for atoms \cite{buijse1989,gritsenko1994,leeuwen1994,leeuwen1995} and diatomics \cite{buijse1989,gritsenko1996a,gritsenko1996b,gritsenko1997}, \red{and to further studies of the conditional wavefunction in the DFT literature \cite{helbig2009,saavedra2016,giorgi2016b,giarrusso2018,kumar2019,kumar2020,giarrusso2020,mccarty2020}.}
  Recently, the formalism of the wavefunction separation has been further developed for the electron-nuclear problem and been termed the exact factorization \cite{abedi2010,abedi2012,gonze2018}.
  The exact factorization has then also been transferred to the many-electron problem as exact electron factorization (EEF) \cite{schild2017,kocak2020,complet}.
  
  In this article we take a closer look at two aspects of the EEF:
  First, in Sec.\ \ref{sec:eef} we present the theory of the EEF and connect it to DFT.
  In the EEF, one electron in the environment of other electrons is described.
  This electron obeys a Schr\"odinger equation with the scalar potential $v$ and a vector potential $\vec{A}$, where $v$ is composed of a part that corresponds to the energy of the environment and a part that relates to the geometric structure of the environment.
  The EEF can be related to DFT by replacing the wavefunction of the interacting electrons that constitute the environment with the corresponding wavefunction of the KS system.
  From the connection of the EEF to DFT, we can interpret $v^{\rm KS}$ and, in particular, $v^{\rm P}$ from the perspective of one electron in the environment of other electrons and, in this way, given them an alternative meaning compared to their standard interpretation.
  Second, in Sec.\ \ref{sec:geopot} we take a closer look at the geometric part $v^{\rm G}$ of $v$.
  We show how $v^{\rm G}$ and $\vec{A}$ are related to the quantum geometric tensor that encodes the geometry of the environment.
  Then, we study how $v^{\rm G}$ behaves for a numerically solvable model of a two-electron homo- and heteronuclear diatomic molecule in one dimension.
  With the help of a two-state model, a quantitative analysis of $v^{\rm G}$ becomes possible and we can show explicitly how changes of the environment are encoded in $v^{\rm G}$.
  Our article closes with a short summary of the findings and ideas for future research.

  \section{Exact electron factorization and density functional theory}
  \label{sec:eef}
  
  In this section, the EEF is presented as a way to reduce an $N$-electron problem to a one-electron problem.
  The resulting one-electron problem is that of one electron in a scalar potential $v$ and a vector potential $\vec{A}$, together representing the environment of the other electrons.
%   Both $v$ and $\vec{A}$ are functionals of an $(N-1)$-electron conditional wavefunction.
  The solution of the one-electron Schr\"odinger equation with these potentials yields the exact one-electron density and current density as well as one-electron observables of the many-electron system.
  
  Next to explaining the physical picture of the EEF, this section has the goal to contrast the usual meaning of the potentials in DFT, i.e., the Hartree-exchange-correlation potential $v^{\rm HXC}$ and the Pauli potential $v^{\rm P}$, with those appearing in the EEF, i.e., the average energy of the $(N-1)$-electron system $v^{\rm H}$ in the presence of an additional electron and the geometric potential $v^{\rm G}$.
  In particular, if the interacting system is replaced with the KS system, $v$ is left unchanged, even though the corresponding $(N-1)$-electron environment that it represents changes.
  This is by construction, as the KS system has the same one-electron density like the interacting system.
%   However, the conditional wavefunction of the KS system is, in general, different from the interacting one, and hence the contributions to $v$ change.
  However, the individual contributions to $v$ change, which can be used relate the DFT potentials $v^{\rm HXC}$ and $v^{\rm P}$ to the EEF potentials $v^{\rm H}$ and $v^{\rm G}$.
  Based on these relations, we show that the idea of one electron being in the environment of the other electrons can be transferred from the EEF to DFT, thus bringing an alternative interpretation to $v^{\rm HXC}$ and, in particular, to $v^{\rm P}$.
  
  \subsection{Formalism of the exact electron factorization}
  
  In the following the Born-Oppenheimer approximation is assumed, i.e., the nuclei are treated as clamped \cite{born1927}.
  For simplicity, we use atomic units and we consider the ground state of a system of $N$ non-relativistic spinless electrons (fermions), i.e., the energetically lowest fully antisymmetric solution $\psi(\vec{r}_1,\dots,\vec{r}_N)$ of a non-relativistic many-electron Hamiltonian for some external potential.
  The generalization to include electron spin and excited states is straightforward but complicates the presentation, hence it is not discussed here.
  For brevity, we sometimes substitute the electronic coordinates with numbers, e.g.\ $\psi(\vec{r}_1,\dots,\vec{r}_N) \equiv \psi(1,\dots,N)$.
  
  The many-electron Schr\"odinger equation is
  \begin{align}
    \left( -\sum_{j=1}^N \frac{\nabla_j^2}{2} + V(1,\dots,N) \right) \psi(1,\dots,N) = E \psi(1,\dots,N)
    \label{eq:se}
  \end{align}
  with the scalar potential 
  \begin{align}
    V(1,\dots,N) = \sum_{j=1}^N v^{\rm ext}(j) + \sum_{j=1}^N \sum_{k=j+1}^N v_{\rm ee}(j,k)
    \label{eq:vel}
  \end{align}
  that is the sum of one-electron (external) potentials $v^{\rm ext}(\vec{r})$ and the electron-electron interaction $v_{\rm ee}(\vec{r}_j,\vec{r}_k)$.
  
\begin{figure*}[!htbp]
    \centering
    \includegraphics[width=.9\linewidth]{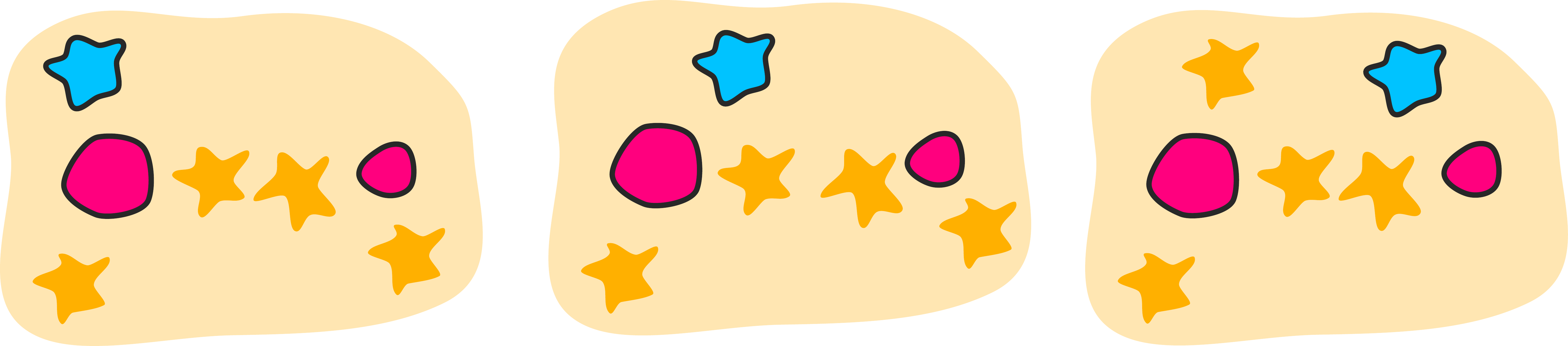}
    \caption{
    Idea of the exact electron factorization illustrated with a diatomic molecule.
    The two round (magenta) shapes represent two nuclei, the five star-like shapes (yellow and blue) represent electrons.
    If the shape has a black border (the nuclei and the blue electron), its position is a condition:
    The nuclei are clamped and the state of the four yellow electrons (the environment) is the conditional wavefunction $\phi(1,2,3,4;\vec{r})$ for a given position $\vec{r}$ of the blue electron.
    The blue electron feels the external potential $v^{\rm ext}(\vec{r})$ that describes the interaction with the nuclei, the potential $v^{\rm H}(\vec{r})$ that is the energy of the 4-electron system given another electron is at $\vec{r}$, and the potential $v^{\rm G}(\vec{r})$ that can be thought of as the additional energy needed to change the state of the 4-electron system when the position $\vec{r}$ of the blue electron is changed.    
    A possible vector potential $\vec{A}(\vec{r})$ might also be felt by the blue electron, e.g., if the system is rotating.
    We emphasize that the blue electron is any electron and the fermionic antisymmetry conditions are unbroken.
    }
    \label{fig:pic_eef_idea}
  \end{figure*}

  The EEF \cite{schild2017} is based on the fact that a joint probability can be written as product of a marginal and a conditional probability \cite{hunter1975,abedi2010}.
  In terms of wavefunctions, this translates to 
  \begin{align}
    \psi(1,\dots,N) = \chi(1) \phi(2,\dots,N;1),
    \label{eq:eef_ansatz}
  \end{align}
  where 
  \begin{align}
    |\chi(\vec{r})|^2 := \braket{\psi(\vec{r},2,\dots,N)|\psi(\vec{r},2,\dots,N)}_{2 \dots N} \equiv \rho(\vec{r}) 
  \end{align}
  is the one-electron density and $\braket{\dots}_{2 \dots N}$ indicates the scalar product (integral) with respect to (w.r.t.) the coordinates $\vec{r}_2, \dots, \vec{r}_N$.
  As the one-electron density is the marginal density of finding an electron at $\vec{r}$ independent of the location of the other electrons, $\chi(\vec{r})$ is called the marginal wavefunction.
  The function 
  \begin{align}
    \phi(2,\dots,N;\vec{r}) := \frac{\psi(\vec{r},2,\dots,N)}{\chi(\vec{r})}
    \label{eq:phi}
  \end{align}
  is the conditional wavefunction whose squared magnitude, $|\phi(2,\dots,N;\vec{r})|^2$,
  represents the conditional probability of finding electrons at 
  $\vec{r}_2, \dots, \vec{r}_N$, given an electron is located at $\vec{r}$.
  Thus, it has to obey the partial normalization condition
  \begin{align}
    \braket{\phi(2,\dots,N;\vec{r}) | \phi(2,\dots,N;\vec{r})}_{2 \dots N} \stackrel{!}{=} 1
    \label{eq:pnc}
  \end{align}
  for all values of $\vec{r}$.
  The conditional wavefunction $\phi(2,\dots,N;\vec{r})$ is the wavefunction of the electrons at $\vec{r}_2, \dots, \vec{r}_N$ under the condition that another electron is at $\vec{r}$. We call the electrons at $\vec{r}_2, \dots, \vec{r}_N$ the environment \cite{kocak2021}.
  The function $\phi$ encodes the spatial electron entanglement \cite{schroeder2017} in the sense that the $N$-electron system is in general not the product of a one-electron wavefunction and an $(N-1)$-electron wavefunction, but that the wavefunction $\phi$ of the $N-1$ electrons depends on where the remaining electron of the $N$-electron system is found (measured).
  From \eqref{eq:eef_ansatz} follows that $\chi(\vec{r})$ obeys 
  the one-electron Schr\"odinger equation\cite{abedi2010,schild2017}
  \begin{align}
    \left( \frac{(-i \prv + \vec{A}(\vec{r}))^2}{2} +  v(\vec{r}) \right) \chi(\vec{r}) &= E \chi(\vec{r})
    \label{eq:eef_chi}
  \end{align}
  with the vector potential
  \begin{align}
    \vec{A}(\vec{r}) &= \braket{\phi(2,\dots,N;\vec{r}) |-i \prv \phi(2,\dots,N;\vec{r})}_{2 \dots N}
  \end{align}
  and with the scalar potential (cf.\ \cite{buijse1989,gritsenko1994,gritsenko1996a,gritsenko1996b})
  \begin{align}
    v(\vec{r}) &= v^{\rm T}(\vec{r}) + v^{\rm V}(\vec{r}) + v^{\rm G}(\vec{r}) + v^{\rm ext}(\vec{r})
    \label{eq:v_eef}
  \end{align}
  that contains the terms
  \begin{align}
    v^{\rm T}(\vec{r})  
      &= \matrixel{\phi(2,\dots,N;\vec{r})}{-\sum\limits_{j=2}^N \frac{\nabla_j^2}{2}}{\phi(2,\dots,N;\vec{r})}_{2 \dots N} 
      \label{eq:eef_vt} \\
    v^{\rm V}(\vec{r})  
      &= \matrixel{\phi(2,\dots,N;\vec{r})}{V(1,2,\dots,N)}{\phi(2,\dots,N;\vec{r})}_{2 \dots N} - v^{\rm ext}(\vec{r})
      \label{eq:eef_vv} \\
    v^{\rm G}(\vec{r}) 
      &= \frac{1}{2} \left( \braket{\prv \phi(2,\dots,N;\vec{r}) | \prv \phi(2,\dots,N;\vec{r})}_{2 \dots N} - \vec{A}(\vec{r})^2 \right).
      \label{eq:eef_vfs}
  \end{align}
  It will be useful for the discussion below to define the sum
  \begin{align}
    v^{\rm H}(\vec{r})   &= v^{\rm T}(\vec{r}) + v^{\rm V}(\vec{r}).   \label{eq:eef_vh}
%     v^{\rm EEF}(\vec{r}) &= v^{\rm H}(\vec{r}) + v^{\rm G}(\vec{r}). \label{eq:eef_veef}
  \end{align}
  The term $v^{\rm T}(\vec{r})$ is the expectation value of the kinetic energy of the  $N-1$ environmental electrons and $v^{\rm V}(\vec{r})$ is the corresponding expectation value of the potential energy, hence $v^{\rm H}(\vec{r})$ is the expectation value of the energy of the environment given one additional electron is at $\vec{r}$.
  The geometric potential $v^{\rm G}(\vec{r})$ is discussed and illustrated in section \ref{sec:geopot}.
  It is connected to how much the conditional wavefunction $\phi$ changes w.r.t.\ $\vec{r}$ and it is needed to calculate the correct kinetic energy of the electron in the presence of the electrons in the environment.
  Together with $\vec{A}(\vec{r})$, it describes the reaction of the environment to an infinitesimal change of the position of the additional electron at $\vec{r}$, and it is related to the Fubini-Study metric as well as to the quantum geometric tensor \cite{provost1980}.
%   We call both potentials $v(\vec{r})$ and $v^{\rm EEF}(\vec{r})$ the EEF potential below, as they only differ by the external potential $v^{\rm ext}(\vec{r})$.
  The meaning of the potentials is explained pictorially in Fig.\ \ref{fig:pic_eef_idea}.
  
  From the product form \eqref{eq:eef_ansatz} the one-electron wavefunction $\chi(\vec{r})$
  is defined only up to a phase, because we can replace $\chi(\vec{r})$ and $\phi(2,\dots,N;\vec{r})$ with
  \begin{subequations}
    \begin{align}
      \tilde{\chi}(\vec{r}) &:= e^{-i S(\vec{r})} \chi(\vec{r})  \\
      \tilde{\phi}(2,\dots,N;\vec{r}) &:= e^{+i S(\vec{r})} \phi(2,\dots,N;\vec{r}),
    \end{align}
    \label{eq:gauge}
  \end{subequations}
  where $S \in \mathbb{R}$, without changing the many-electron wavefunction $\psi(1,\dots,N)$ and without violating the partial normalization condition \eqref{eq:pnc}.
  The equations for $\chi$ (equation \eqref{eq:eef_chi}) and $\phi$ (see the Supplemental 
  Material of \cite{schild2017}) also do not change under \eqref{eq:gauge} if $\vec{A}(\vec{r})$ is replaced with 
  \begin{align}
    \tilde{\vec{A}}(\vec{r}) = \vec{A}(\vec{r}) + \prv S(\vec{r}).
    \label{eq:Agauge}
  \end{align}
  The choice of $S$ is thus arbitrary, it is a gauge freedom of the theory.
  The measurable quantities of the theory need to be gauge invariant, i.e., they cannot depend on the choice of $S$.
  The potentials $v^{\rm T}$, $v^{\rm V}$, and $v^{\rm G}$ have this property, as shown in Appendix \ref{sec:eefpot}.
  Also, 
  \begin{align}
    \hat{p} = -i \prv + \vec{A}(\vec{r})
  \end{align}
  is the gauge-invariant canonical momentum  and 
  \begin{align}
    \hat{T} = \frac{(-i \prv + \vec{A}(\vec{r}))^2}{2} + v^{\rm G}(\vec{r})
    \label{eq:ekin_gi}
  \end{align}
  is the gauge-invariant kinetic energy of an electron in the environment of the other electrons.
  
  An important feature of the EEF is that the many-electron problem is replaced with the one-electron problem \eqref{eq:eef_chi} for the one-electron wavefunction $\chi$.
  If the (components of the) potentials $v$ and $\vec{A}$ were known, one-electron observables of the many-electron system (like the dipole or momentum) could be directly calculated from $\chi$ and the energy of the many-electron system $E$ could be obtained.
  However, the conditional wavefunction $\phi$ is needed to obtain these potentials, and its determining equation is difficult to solve exactly \cite{gossel2019}.
  Nevertheless, the EEF formalism provides explicit expressions for the needed one-electron potentials in terms of $\phi$ (or in terms of the full many-electron wavefunction $\psi$) that can be used to find suitable approximations or to connect to DFT.
  
  \subsection{Relation to density functional theory}
  
  We can relate the EEF to DFT by introducing the KS system.
  In KS-DFT \cite{kohn1965}, the interacting many-electron system is replaced by a non-interacting many-electron system, the KS system, with the same one-electron density as for the interacting problem.
  The wavefunction of the KS system is
  \begin{align}
    \psi^{\rm KS}(1,\dots,N)
      &=  \hat{A} \left( \prod_{j=1}^N \vp_j^{\rm KS}(j) \right) 
      \label{eq:psiks}
  \end{align}
  where $\hat{A}$ is an anti-symmetrization operator and we require $\Braket{\psi^{\rm KS} | \psi^{\rm KS}} = 1$.
  The KS orbitals $\vp_j^{\rm KS}(\vec{r})$ are obtained by solving the one-electron Schr\"odinger equation 
  \begin{align}
    \left( -\frac{\prv^2}{2} + v^{\rm KS}(\vec{r}) \right) \vp_j^{\rm KS}(\vec{r}) = \varepsilon_j^{\rm KS} \vp_j^{\rm KS}(\vec{r}).
  \end{align}
  with the KS potential
  \begin{align}
    v^{\rm KS}(\vec{r}) = v^{\rm HXC}(\vec{r}) + v^{\rm ext}(\vec{r}).
  \end{align}
  The potential $v^{\rm HXC}$ is the Hartree-exchange-correlation potential.
  We do not separate it further in the following discussion.
  The one-electron density of the interacting system is
  \begin{align}
    \rho(\vec{r}) 
      \equiv \frac{1}{N} \sum_{j=1}^N |\vp_j^{\rm KS}(\vec{r})|^2.
      \label{eq:norm}
  \end{align}
  In contrast to the usual convention of normalizing the one-electron density to the number of electrons, we require that $\braket{\psi|\psi} = \braket{\rho(\vec{r})} = \braket{|\vp_j^{\rm KS}(\vec{r})|^2} = 1$, i.e., the many-electron wavefunction, the one-electron density, and the KS orbitals are all normalized to 1.
  We make this non-standard choice for $\rho$ because we interpret the density $\rho(\vec{r})$ as the density of one electron in the environment of the other electrons, as discussed below.
  While the KS system exists and is unique \cite{hohenberg1964}, the KS potential $v^{\rm KS}$ cannot be directly obtained from the many-electron wavefunction $\psi$ and/or the one-electron density $\rho$.
  Different numerical methods exist to find the exact $v^{\rm KS}$ for a given one-electron density, and some recent discussions and applications  of this inverted KS problem can be found in \cite{kananenka2013,jensen2018,kumar2019,callow2020}.
  
  The EEF equation \eqref{eq:eef_chi} is equivalent to the central equation 
  of OF-DFT,
  \begin{align}
    \left( -\frac{\prv^2}{2} + v^{\rm KS}(\vec{r}) + v^{\rm P}(\vec{r}) \right) \sqrt{\rho(\vec{r})} &= \mu \sqrt{\rho(\vec{r})}
    \label{eq:ofdft}
  \end{align}
  where $\mu = \varepsilon_N^{\rm KS}$ is the chemical potential (the eigenvalue of the highest occupied KS orbital) and $v^{\rm P}$ is the Pauli potential.
  From \eqref{eq:pauli} we see that \eqref{eq:ofdft} is identical to the EEF equation \eqref{eq:eef_chi} if we fix the gauge as $\vec{A}(\vec{r}) = 0$
  and if $\chi(\vec{r}) = \sqrt{\rho(\vec{r})}$, i.e., if $\chi$ is real-valued.
  This gauge choice cannot always be made \cite{requist2016,requist2017} but is supposed to be possible for the (non-degenerate) ground state of the many-electron system with zero total angular momentum and possibly also for other states without total angular momentum.\footnote{
    The gauge $\vec{A}(\vec{r}) = 0$ implies that the vector potential is curl-free in any gauge. 
    This condition may be violated even if the total angular momentum vanishes, hence we cannot make a definite statement here.}
  Then, the EEF potential is related to KS and Pauli potentials as
  \begin{align}
    v(\vec{r}) &= v^{\rm KS}(\vec{r}) + v^{\rm P}(\vec{r}) \\
     v^{\rm H}(\vec{r}) + v^{\rm G}(\vec{r}) &= v^{\rm HXC}(\vec{r}) + v^{\rm P}(\vec{r})
  \end{align}
  up to a constant $E - \mu$ to be added on the right-hand side of the equations.
  
  The Pauli potential can be written in terms of the KS system as\cite{levy1988,kraisler2020}
  \begin{align}
    v^{\rm P}(\vec{r}) 
      = v^{\rm PH}(\vec{r}) + v^{\rm PG}(\vec{r}) 
    \label{eq:pauli_ks}
  \end{align}
  with 
  \begin{align}
    v^{\rm PH}(\vec{r}) 
      &= \sum_{n=1}^N (\varepsilon_{N}^{\rm KS} - \varepsilon_n^{\rm KS}) |\phi_n^{\rm KS}(\vec{r})|^2 
      \label{eq:ph} \\
    v^{\rm PG}(\vec{r}) 
      &= \frac{1}{2} \sum_{n=1}^N |\prv \phi_n^{\rm KS}(\vec{r})|^2,
      \label{eq:pg}
  \end{align}
  where we used the functions
  \begin{align}
    \phi_n^{\rm KS}(\vec{r}) = \frac{\vp_n^{\rm KS}(\vec{r})}{\sqrt{\rho(\vec{r})}}.
    \label{eq:phi_n_KS}
  \end{align}
  
  We now relate the EEF formalism to DFT by realizing that the functions \eqref{eq:phi_n_KS} are similar to the conditional wavefunction $\phi$ of the EEF and may be interpreted as KS orbitals of the environment.
  In particular, we define the conditional wavefunction 
  \begin{align}
    \phi^{\rm KS}(2,\dots,N;\vec{r}) = \frac{\psi^{\rm KS}(\vec{r},2,\dots,N)}{\sqrt{\rho(\vec{r})}}
    \label{eq:kscond}
  \end{align}
  of the non-interacting KS system, where $\psi^{\rm KS}$ is given by \eqref{eq:psiks}, and we can interpret the potential $v$ of as functional $v = v[\phi,V]$ of the conditional wavefunction $\phi$ and the many-electron potential $V(1,\dots,N)$, see \eqref{eq:v_eef}-\eqref{eq:eef_vfs}.
  As both $\psi$ and $\psi^{\rm KS}$ correspond to the same one-electron density $\rho(\vec{r})$, it follows that 
  \begin{align}
    v[\phi,V] = v[\phi^{\rm KS},V^{\rm KS}], 
    \label{eq:v_vs_vni}
  \end{align}
  where $V^{\rm KS}(1,\dots,N) = \sum_{j=1}^N v^{\rm KS}(j)$ is the many-electron potential of the non-interacting KS system.
  Relation \eqref{eq:v_vs_vni} states that the same one-electron potential $v$ is obtained if it is evaluated as functional with the exact quantities or with the KS quantities.
  
  If we also interpret $v^{\rm H}$ and $v^{\rm G}$ as functionals $v^{\rm H} = v^{\rm H}[\phi,V]$ and $v^{\rm G} = v^{\rm G}[\phi]$, we find (cf.\ \cite{gritsenko1996b})
  \begin{subequations}
    \begin{align}
      v^{\rm PH}(\vec{r}) + v^{\rm HXC}(\vec{r})
        &= v^{\rm H}[\phi^{\rm KS},V^{\rm KS}](\vec{r}) \label{eq:vh_corr}  \\
      v^{\rm PG}(\vec{r}) 
        &= v^{\rm G}[\phi^{\rm KS}](\vec{r}), \label{eq:vg_corr}
    \end{align}
    \label{eq:v_corr}
  \end{subequations}
  where \eqref{eq:vh_corr} holds up to a constant due to the different asymptotic conditions in DFT and in the EEF, as explained above.
  We thus see that $v^{\rm PG}(\vec{r})$ is the geometric potential of the $N-1$ non-interacting electrons of the KS system if one additional electron is at $\vec{r}$.
  Also, $v^{\rm H}[\phi^{\rm KS},V^{\rm KS}]$ is the corresponding energy of those $N-1$ electrons. 
%   That $v^{\rm HXC}$ appears in \eqref{eq:vh_corr} is because the role of the external potential $v^{\rm ext}(\vec{r})$ that the electron in the interacting system feels is played by $v^{\rm KS}(\vec{r})$ for the electron in the KS system.
  The left-hand side and the right-hand side of \eqref{eq:vh_corr} are only equal up to a constant (which is $\sum_{j=1}^{N-1} \varepsilon_j^{\rm KS}$), because $v$ in \eqref{eq:eef_chi} and $v^{\rm KS} + v^{\rm P}$ in \eqref{eq:ofdft} are shifted relative to each other.
  In the EEF, the asymptotic value  $\lim_{|\vec{r}| \rightarrow \infty} v(\vec{r})$ is the energy of the ionized system, while in OF-DFT the potential typically is shifted such that it becomes zero for large $|\vec{r}|$.
  
  The reason why we can work with the conditional KS orbitals $\phi_n^{\rm KS}(\vec{r})$ instead of the full conditional KS-wavefunction $\phi^{\rm KS}(\vec{r})$ to obtain $v^{\rm PH}$ and $v^{\rm PG}$ is the orthogonality of the KS-orbitals w.r.t.\ integration over the electronic coordinates. 
  This orthogonality can be used to simplify the expressions for the potentials $v^{\rm PH}$ and $v^{\rm PG}$ such that no integration is left in \eqref{eq:ph} and \eqref{eq:pg}.
  
  \subsection{Interpreting the DFT potentials}
  
  Via \eqref{eq:v_corr}, the EEF provides a different view on the Hartree-exchange-correlation potential $v^{\rm HXC}$ and, in particular, on the Pauli potential $v^{\rm P}$.
  When OF-DFT and KS-DFT are compared, a central point of discussion is how the two theories treat the fermionic antisymmetry of a many-electron system.
  The symmetry constraints for the many-electron wavefunction $\psi(1,\dots,N)$ are included in an elegant way in KS-DFT via the construction of the non-interacting KS system, which has the same one-electron density like the interacting system, but which also corresponds to an antisymmetric many-electron wavefunction $\psi^{\rm KS}(1,\dots,N)$.
  OF-DFT, however, is sometimes interpreted as mapping to a non-interacting bosonic system with the same one-body density.
  The Pauli potential is thus often viewed as necessary to describe the antisymmetry correctly, because it is the difference potential between the supposed fermionic and bosonic non-interacting systems. \cite{march1986,levy1988}.
  
  While the construction of the non-interacting bosonic system is technically correct, the EEF provides another interpretation:
%   It splits the joint $N$-electron wavefunction into a product of a one-electron wavefunction $\chi$ and the $(N-1)$-electron wavefunction $\phi$ of the environment.
  Despite the product form \eqref{eq:eef_ansatz}, the fermionic antisymmetry constraints are unbroken:
  The wavefunction $\phi$ fulfills the symmetry constraints w.r.t.\ exchange of the (spin- \& spatial) coordinates of the electrons in the environment.
  The antisymmetry constraints w.r.t.\ the additional electron are found in the product $\chi(1) \phi(2,\dots,N;1)$ and are, thus, implicitly contained in the EEF formalism.
  
  In the EEF picture, there is thus no Pauli potential which turns a (non-interacting) bosonic system into a  fermionic system, but the interacting fermionic system itself is considered from the start.
  The EEF potentials $v^{\rm H}$ and $v^{\rm G}$ have a clear physical meaning in terms of how one electron feels the environment provided by the other electrons:
  $v^{\rm H}$ is the energy of the other electrons and $v^{\rm G}$ is an additional resistance that the electron experiences if its change of position leads to a change of the state of the other electrons (i.e., if there is a strong spatial entanglement).
  
  If the interacting system is replaced with the KS system, we see from \eqref{eq:v_corr} that the corresponding geometric potential becomes one part of the Pauli potential, while the energy of the environment becomes the sum of $v^{\rm HXC}$ with the other part of the Pauli potential.
  Moreover, by evaluating \eqref{eq:vh_corr} explicitly, we have (up to a constant)
  \begin{align}
    v^{\rm H}[\phi^{\rm KS},V^{\rm KS}] + v^{\rm ext}
      &= \Braket{\phi^{\rm KS} | -\sum_{j=2}^N \frac{\nabla_j^2}{2} + \sum_{j=1}^N v^{\rm KS}(j) | \phi^{\rm KS}}_{2 \dots N} \\
      &= v^{\rm PH} + \underbrace{\Braket{\phi^{\rm KS} | v^{\rm KS}(1) | \phi^{\rm KS}}_{2 \dots N}}_{v^{\rm HXC} + v^{\rm ext}}.
  \end{align}
  Thus, we can actually think of the Pauli potential as being the EEF potential for the KS system, with $v^{\rm HXC}$ being a correction due to the different way of how the electron-electron interaction is described.
  
  In contrast to the usual view on DFT and especially on OF-DFT, one does, from the EEF perspective, not talk about a fermionic or a bosonic many-electron problem.
  Instead, the problem of one electron in the environment of other electrons is considered both for the interacting many-electron problem (the EEF) and for the KS system (OF-DFT).
  In both cases the same one-electron potential $v$ is obtained, but because the environment is described differently, the contributions to $v$ can differ.

  \section{The geometric potential}
  \label{sec:geopot}
  
  While the average energy $v^{\rm H}$ of the environment is straightforward to understand in the EEF, the meaning of the geometric potential $v^{\rm G}$ is less obvious.
  As \eqref{eq:ekin_gi} shows, $v^{\rm G}$ is one part of the gauge-invariant kinetic energy operator $\hat{T}$ and $\braket{\chi|\hat{T}|\chi}$ is the expectation value of the kinetic energy of one electron in the environment of the other $N-1$ electrons.
  Also, in appendix \ref{sec:ekind} we show that $v^{\rm G}$ is related to the kinetic energy density (cf.\ \cite{buijse1989}) and that the different kinetic energy densities of the interacting and the KS system fully account for the differences between $v^{\rm G}$ and $v^{\rm PG}$.
  Hence, $v^{\rm G}$ has been called the ``kinetic potential'' in the literature \cite{gritsenko1994,giarrusso2020}.
  
  However, there is a geometric meaning attached to $v^{\rm G}$ and, in connection to this, also to the vector potential $\vec{A}$.
  In this section, we first show how $v^{\rm G}$ and $\vec{A}$ are related to the quantum-geometric tensor that describes the geometric structure of the environment of the one-electron system.
  We then proceed by examining $v^{\rm G}$ for numerically solvable models of a homonuclear and a heteronuclear diatomic in one-dimension.
  Finally, we connect to the geometric picture of $v^{\rm G}$ by investigating how the dependence of the environment on the one electron are encoded in $v^{\rm G}$ via a two-state model.
%   
%   As the EEF scalar potential $v^{\rm EEF}$ is the sum of $v^{\rm HXC}$ and the Pauli potential $v^{\rm P}$, there arises the question how $v^{\rm PH} + v^{\rm HXC}$ and $v^{\rm PG}$ for the non-interacting KS system differ from their counterparts $v^{\rm H}$ and $v^{\rm G}$ for the interacting system.
%   From \eqref{eq:v_corr} it is clear that these functions need not be identical because $\phi^{\rm KS}$ and $V^{\rm KS}$ in general differ from $\phi$ and $V$.
%   Yet, \eqref{eq:v_vs_vni} shows that there is a strong connection between them.
%   This is investigated in the following with the help of simple numerically solvable models.
%   
  
  All model systems described in the following were solved numerically with the program package QMstunfti \cite{qmstunfti} that is based on the sparse-matrix functionality of Scipy \cite{scipy}, which in turn partially uses the ARPACK library \cite{arpack}.
  The exact KS potentials were obtained from the inversion procedure used in \cite{hodgson2017}, see also \cite{leeuwen1994}.
  We choose the gauge of zero vector potential, which is always possible for one-dimensional finite systems.
  
  \subsection{The quantum-geometric tensor}
  \label{sec:qgt}
  
  To better understand $v^{\rm G}$, we describe its relation to the quantum geometric tensor.
  We consider a general function $f(x;t)$ that is an element of a Hilbert space with inner product defined w.r.t.\ the coordinate(s) $x$.
  The function $f$ has an additional dependence on a parameter $t$, and it shall also be normalized as $\braket{f(x;t)|f(x;t)}_x = 1$.
  We are interested in the change of $f$ with $t$.
  Taking the norm of the difference between $f(x;t)$ and $f(x;t+dt)$,
  \begin{align}
    df^2 = || f(x;t+dt) - f(x;t) ||^2 = \Braket{f(x;t+dt) - f(x;t) | f(x;t+dt) - f(x;t)}_x
  \end{align}
 we find for infinitesimal $dt$ that
 \begin{align}
   df^2 = \Braket{\pt f(x;t) | \pt f(x;t)}_x dt^2.
 \end{align}
 The term
 \begin{align}
    \tilde{g} = \Braket{\pt f(x;t) | \pt f(x;t)}_x \ge 0
    \label{eq:g01}
 \end{align}
 looks like a metric that represents how $f$ changes when $t$ is changed.
 
 However, \eqref{eq:g01} is ambiguous, as $f(x;t)$ is only defined up to an $x$-independent phase.
 In particular, we can define
 \begin{align}
   f(x;t) \rightarrow f(x;t) e^{-i S(t)}
 \end{align}
 with $S \in \mathbb{R}$ without changing the state that $f$ represents.
 The choice of $S(t)$ is a gauge choice and, for $\tilde{g}$ to be a useful metric, it should be independent of the choice of gauge.
 This is achieved by replacing \eqref{eq:g01} with the Fubini-Study metric \cite{provost1980,abe1993}
 \begin{align}
    g = \Braket{\pt f(x;t) | P_{\perp} |  \pt f(x;t)}_x \ge 0
    \label{eq:g02}
 \end{align}
  where $P_{\perp}$ is a projector into the space orthogonal to $f$, 
  \begin{align}
    P_{\perp} = 1 - \ket{f(x;t)} \bra{f(x;t)}_x.
  \end{align}
  
  If $f$ depends on multiple parameters, $f(x;t_1,\dots,t_n)$, the change of $f$ with the parameters is described by the quantum-geometric tensor \cite{provost1980,berry1989}
  \begin{align}
    T_{ij} = \Braket{\partial_{t_i} f(x;t_1,\dots,t_n) | P_{\perp} | \partial_{t_j} f(x;t_1,\dots,t_n)}.
  \end{align}
  This tensor yields the Fubini-Study metric tensor as its real part,
  \begin{align}
    g_{ij} = \frac{1}{2}(T_{ij} + T_{ji}), 
  \end{align}
  and its imaginary part is the Berry curvature \cite{berry1984}
  \begin{align}
    B_{ij} = \frac{1}{2i}(T_{ij} - T_{ji}).
  \end{align}
  
  We can compare these definitions to the geometric potential $v^{\rm G}$.
  There, $f(x;t) \rightarrow \phi(2,\dots,N;\vec{r})$ where the coordinates $\vec{r}$ of one electron are parameters.
  As we restrict the discussion to a simple kinetic energy operator \eqref{eq:se} with Cartesian coordinates for the particles, there are no terms coupling the three components $r_i$ of $\vec{r}$ and, thus, the corresponding Fubini-Study metric tensor is diagonal with components
  \begin{align}
    g_{ii} = \Braket{ \partial_{r_i} \phi(2,\dots,N;\vec{r}) | P_{\perp} | \partial_{r_i} \phi(2,\dots,N;\vec{r})}_{2 \dots N}
  \end{align}
  The metric is 
  \begin{align}
    ds^2 = \sum_i \sum_j g_{ij} dr_i dr_j
    \label{eq:ds01}
  \end{align}
  and the geometric potential is
  \begin{align}
   v^{\rm G}(\vec{r}) = \frac{\hbar^2}{2 m_{\rm e}} ds^2,
  \end{align}
  where we added Plank's constant and the electronic mass for emphasis.
  In atomic units and for the simple kinetic energy term in \eqref{eq:se}, the geometric potential is 
  \begin{align}
    v^{\rm G}(\vec{r}) = \frac{1}{2} \sum_{i=1}^3 g_{ii}.
  \end{align}
  It measures how much the wavefunction $\phi$ of the electrons in the environment changes when the position $\vec{r}$ of the conditional electron is changed.
  As it is a distance measure, $v^{\rm G}(\vec{r}) \ge 0$, which is also obvious from its definition.
  Interpreted as potential, it repels the electron from regions where the environment changes significantly along $\vec{r}$.
  
  Next to the metric tensor, the Berry curvature also contains important information about the geometry of the problem.
  In particular, its gauge-invariant components can be expressed in terms of the components of the vector potential $\vec{A}(\vec{r})$ as 
  \begin{align}
    B_{ij} = \partial_{r_i} A_j - \partial_{r_j} A_i.
  \end{align}
  It can be shown (e.g.\ by noting that $B_{ij}$ is the exterior derivative of $\vec{A}$ and by using the theory of differential forms \cite{flanders1989}) that, if any component $B_{ij} \ne 0$, the vector potential $\vec{A}(\vec{r})$ cannot be written as gradient of a scalar field, $\vec{A}(\vec{r}) \ne \nabla_{\vec{r}} S(\vec{r})$ for any $S \in \mathbb{R}$.
  In this case, we see from a comparison to \eqref{eq:Agauge} that the choice of gauge $\vec{A}(\vec{r}) \stackrel{!}{=} 0$ cannot be made.
  
  Below, we only consider (finite) one-dimensional systems for which $\vec{A}$ is a scalar field that can always be written as a gradient field.
  Then, the choice $A \stackrel{!}{=} 0 $ is possible.
  We make this choice and do not consider the vector potential or Berry curvature further.
  However, these quantities might have to be taken into account for the study of the full three-dimensional problem.
  Current results suggest that the choice $\vec{A}(\vec{r}) \stackrel{!}{=} 0$ is possible for the three-dimensional problem when the system is not rotating \cite{requist2016}, but further investigations are needed to draw a definite conclusion.
  
  \subsection{Model study of a one-dimensional homonuclear diatomic molecule}
  
  As example for the behavior of $v^{\rm G}$, we consider a one-dimensional model of a diatomic molecule with two electrons and look at the lowest antisymmetric state.
  This state corresponds to the triplet state is spin was included.
  We select this state because for the symmetric (singlet) ground-state there is no difference between the EEF and DFT, as only one DFT orbital is occupied.
  
  The Hamiltonian for clamped nuclei is 
  \begin{align}
    H_{\rm m}^{(Z)}(R) = \sum_{j=1}^2 \left(-\frac{\partial_j^2}{2} + v_{\rm en}(x_j;R,Z) \right) + v_{\rm ee}(x_1,x_2) + v_{\rm nn}(R,Z),
    \label{eq:h_mol}
  \end{align}
  where the electron-nuclear interaction is given by
  \begin{align}
    v_{\rm en}(x;R,Z) = -\frac{Z}{\sqrt{ (x+R/2)^2 + c_{\rm en} }} -\frac{1}{\sqrt{ (x-R/2)^2 + c_{\rm en} }},
  \end{align}
  corresponding to two nuclei with charges $Z$ and $+1$ located at $-R/2$ and $R/2$, respectively.
  The electron interaction is 
  \begin{align}
    v_{\rm ee}(x_1,x_2) = \frac{1}{\sqrt{ (x_1-x_2)^2 + c_{\rm ee} }}
    \label{eq:vee}
  \end{align}
  and the nuclear interaction is given by
  \begin{align}
    v_{\rm nn}(R,Z) = \frac{Z}{\sqrt{R^2 + c_{\rm nn}}}.
  \end{align}
  The external potential for this model is 
  \begin{align}
    v^{\rm ext}(x) &= v_{\rm en}(x;R,Z) + v_{\rm nn}(R,Z).
  \end{align}
  The parameters are \unit[$c_{\rm en} = c_{\rm ee} = 0.5$]{$a_0^2$} and \unit[$c_{\rm nn} = 0.1$]{$a_0^2$}.
  We first consider the homonuclear (symmetric) case with $Z=+1$.
  
  \begin{figure}[!htbp]
    \centering
    \includegraphics[width=.99\linewidth]{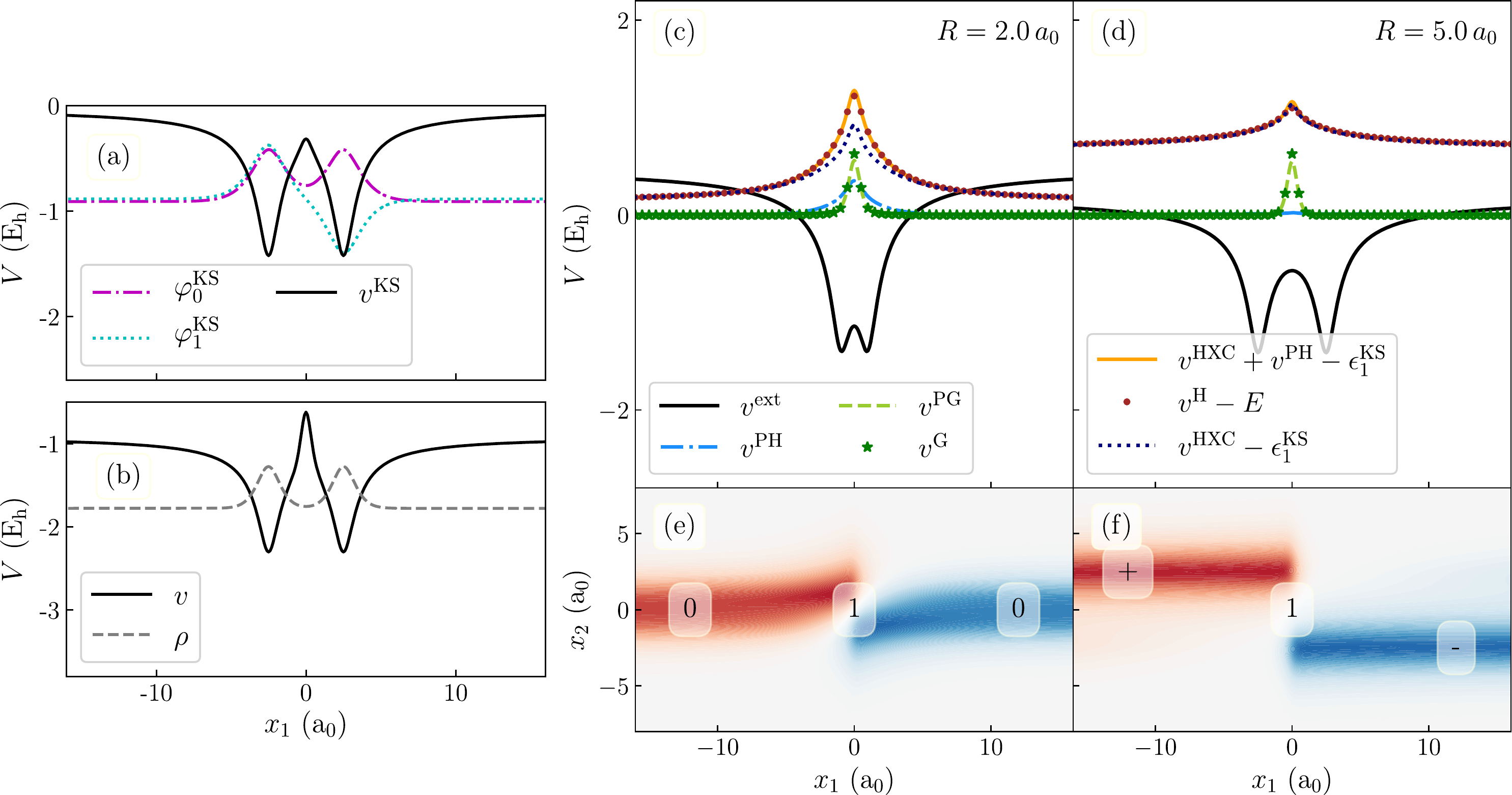}
    \caption{
    (a) KS potential $v^{\rm KS}$ and occupied KS orbitals $\varphi_j^{\rm KS}$ (shifted to their corresponding eigenvalues $\varepsilon_j^{\rm KS}$) as well as (b) EEF/OF-DFT potential $v$ and one-electron density $\rho$ (shifted to its energy) for the energetically lowest antisymmetric electronic state of the homonuclear two-electron diatomic molecule with internuclear distance \unit[$R=5$]{$a_0$}.
    $v^{\rm KS}$ is shifted such that it is zero for large $|x_1|$, while the limit $|x_1| \rightarrow \infty$ of $v$ is the ground-state energy of the ionized system.
    (c), (d):
    Contributions to $v$ in the EEF and in DFT for two different internuclear distances $R$ of the homonuclear two-electron diatomic molecule, for its energetically lowest antisymmetric electronic state. 
    (e), (f): 
    Conditional wavefunctions $\phi(x_2;x_1)$ corresponding to the panels above, shown as contour plots (color indicates the sign).
    The states discussed in Sec.\ \ref{sec:twostatemodel} in the context of the geometric potential $v^{\rm G}$ are marked as ``0'', ``1'', ``+'', and ``-''.
    }
    \label{fig:pic_geo_sym}
  \end{figure}
  
%   Fig.\ \ref{fig:pic_potcomp_sym} shows the EEF potential together with the one-electron density as well as the exact KS potential together with the two lowest KS-orbitals for an internuclear distance of \unit[$R = 5$]{$a_0$}.
%   For this case and , the KS potential is a symmetric double-well potential and the KS-orbitals are almost degenerate and look like typical tunneling states of a double well potential, with the wavefunction of the energetically lower state having the same sign in both wells, while the wavefunction of the energetically higher state switches sign at $x_1=0$.
%   The EEF potential and the KS potential look similar, but the EEF potential has a higher barrier at $x_1=0$.

  Fig.\ \ref{fig:pic_geo_sym}a shows the exact KS potential of the model together with the two lowest KS-orbitals for an internuclear distance of \unit[$R = 5$]{$a_0$}, and Fig.\ \ref{fig:pic_geo_sym}b shows the EEF potential together with the one-electron density.
  The KS-orbitals are almost degenerate and look like typical tunneling states of a double well potential, with the wavefunction of the energetically lower state having the same sign in both wells, while the wavefunction of the energetically higher state switches sign at $x_1=0$.
  
  The components of the EEF potential $v$ and the DFT potentials are depicted in Fig.\ \ref{fig:pic_geo_sym}c and \ref{fig:pic_geo_sym}d for the internuclear distances \unit[$R = 2$]{$a_0$} and \unit[$R = 5$]{$a_0$}, respectively.
  For \unit[$R = 2$]{$a_0$} the two electrons are relatively close to each other.
  All parts of $v$ except the external potential $v^{\rm ext}$ are repulsive bell-shaped potentials centered around $x_1 = 0$.
  
  For the chosen parameters, the model is well-described by Hartree-Fock theory and the KS wavefunction is very close to the interacting wavefunction.
  Thus, the parts of $v$ based on the KS quantities and in the EEF are similar,
  \begin{subequations}
    \begin{align}
      v^{\rm PG} &\approx v^{\rm G} \\
      v^{\rm PH} + v^{\rm HXC} &\approx v^{\rm H}
    \end{align}
    \label{eq:pksimfs}
  \end{subequations} 
  with the second relation holding up to a constant.
  However, some differences of up to \unit[0.08]{$E_h$} exists, with  $v^{\rm PG}$ being slightly smaller than $v^{\rm G}$ and $v^{\rm PH} + v^{\rm HXC} - \varepsilon_1^{\rm KS}$ being slightly larger than $v^{\rm H} - E$.
  
  When the interatomic distance $R$ is increased, the system becomes two separated one-electron atoms and the electron-electron interaction decreases.
  Consequently, $v^{\rm HXC}$ rapidly becomes zero with increasing $R$ and $v^{\rm PH} \approx v^{\rm H}$.
  The energy $v^{\rm H}$ (or $v^{\rm PH}$) of the environment becomes more and more a constant shift to the potential: 
  It is the energetic contribution of the electron in the environment (at $x_2$) to the potential felt by the electron at $x_1$, and the wavefunction $\phi(x_2;x_1)$ of the environment is approximately the ground state of one of the separated atoms.
  This is visible in $\phi(x_2;x_1)$, shown in Fig.\ \ref{fig:pic_geo_sym}e and \ref{fig:pic_geo_sym}f:
  As $\phi(x_2;x_1)$ is the wavefunction of one electron at $x_2$ given there is another one at $x_1$, we have that for a large internuclear distance $R$ the electron at $x_2$ is either located at one nucleus or at the other.
  For \unit[$R=5$]{$a_0$}, given one electron of the two-electron system is found at, say, \unit[$x_1<-2$]{$a_0$}, it is likely to ``originate'' from the nucleus at \unit[$x_1=-2.5$]{$a_0$}.
  The second electron is thus most likely found at the other nucleus centered around \unit[$x_2=+2.5$]{$a_0$} and $\phi(x_2;x_1)$ corresponds approximately to the ground state of an electron at that nucleus.
%   Similarly, if one electron of the two-electron system is found on the opposite side at $x_1>+2$, the conditional wavefunction $\phi(x_2;x_1)$ of the second electron is approximately that of the ground state of the atom centered around \unit[$x_2=-4$]{$a_0$}. 
  
  Turning to the geometric potential $v^{\rm G}$, we observe what is expected from the discussion in section \ref{sec:qgt}:
  The potential $v^{\rm G}$ is a measure of how strong the wavefunction $\phi(x_2;x_1)$  of the one-electron environment changes if the other electron is moved along $x_1$.
  \red{There is a large change in the conditional wavefunction only at $x_1 \approx 0$ that is reflected in $v^{\rm G}$ as a peak in this region (see also \cite{gritsenko1996b,giarrusso2018} for such a peak in similar models).}
  If the internuclear distance is increased, the system becomes more and more that of two separated atoms and the change of the conditional wavefunction at $x_1 \approx 0$ becomes sharper.
  With increasing internuclear distance the peak of the geometric potential $v^{\rm G}$ (or $v^{\rm PG}$) then becomes somewhat more localized at $x_1=0$ and also slightly higher, although this is hardly visible in the figure.
  We emphasize that the shape of $v^{\rm G}$ has nothing to do with the sign change of $\phi(x_2;x_1)$ along $x_1$, as can been seen from the definition \eqref{eq:eef_vfs} of $v^{\rm G}$.
  It is for example also present in the symmetric ground state of $H_{\rm m}^{(1)}$ for which no sign change in $\phi(x_2;x_1)$ happens but $\phi(x_2;x_1)$ is otherwise similar (cf.\ \cite{buijse1989,benitez2016})
  
  \subsection{Model study of a one-dimensional heteronuclear diatomic molecule}
  
  We now investigate what changes if we have a heteronuclear diatomic molecule instead of a homonuclear one.
  For this purpose, we use $Z=2$ in the Hamiltonian \eqref{eq:h_mol}, such that there is one nucleus at $+R/2$ with charge $+1$ and one nucleus at $-R/2$ with charge $+2$, and we again consider the lowest antisymmetric state.
  
  \begin{figure}[!htbp]
    \centering
    \includegraphics[width=.99\linewidth]{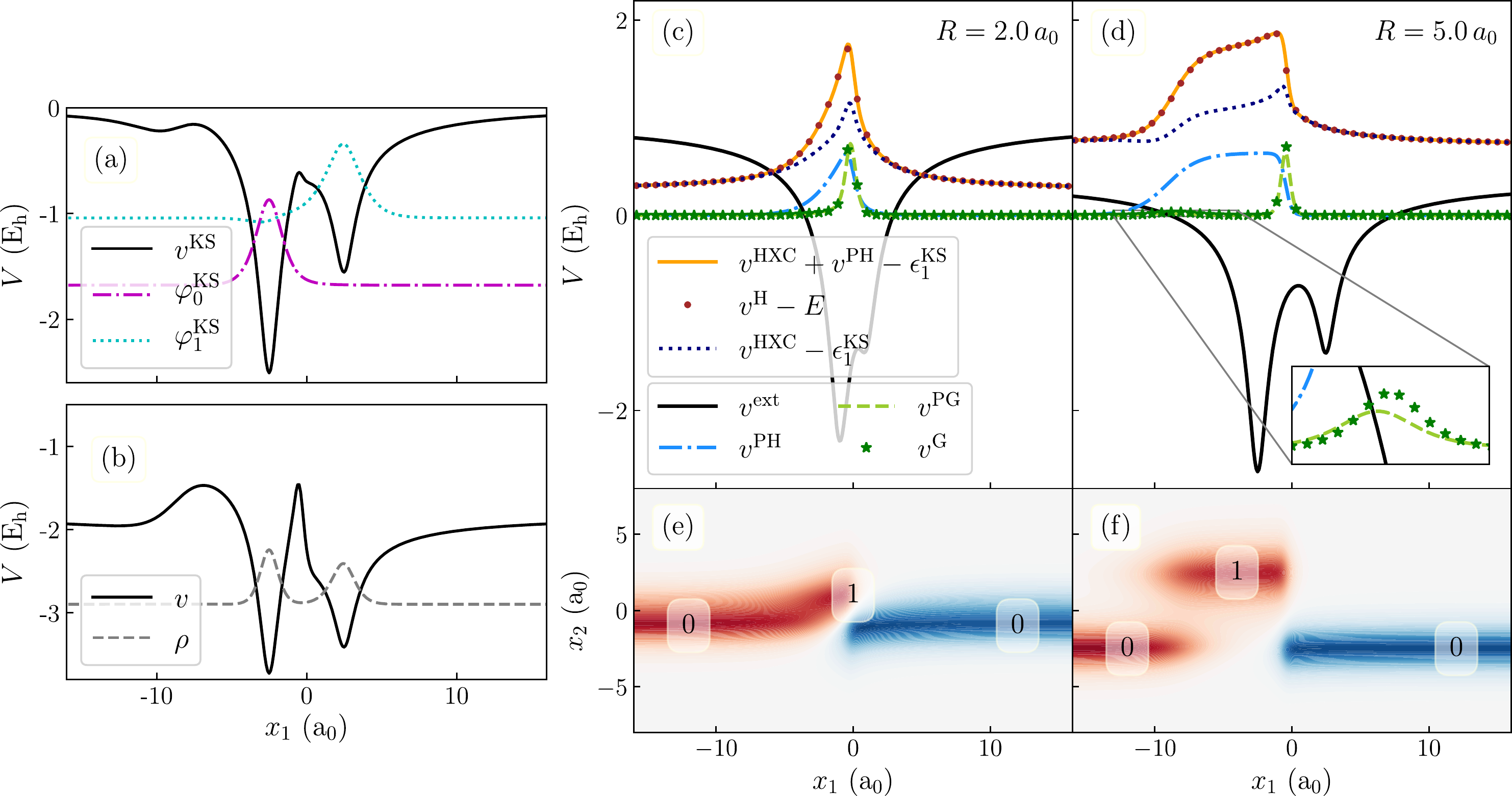}
    \caption{
    (a) KS potential $v^{\rm KS}$ and occupied KS orbitals $\varphi_j^{\rm KS}$ (shifted to their corresponding eigenvalues $\varepsilon_j^{\rm KS}$) as well as (b) EEF/OF-DFT potential $v$ and one-electron density $\rho$ (shifted to its energy) for the energetically lowest antisymmetric electronic state of the heteronuclear two-electron diatomic molecule with internuclear distance \unit[$R=5$]{$a_0$}.
    $v^{\rm KS}$ is shifted such that it is zero for large $|x_1|$, while the limit of $|x_1| \rightarrow \infty$ of $v$ is the ground-state energy of the ionized system.
    (c), (d): 
    Contributions to $v$ in the EEF and based on KS quantities for two different internuclear distances $R$ of the heteronuclear two-electron diatomic molecule, for the energetically lowest antisymmetric electronic state. 
    In (d), an inset shows the details of $v^{\rm geo}$ for $x_1 \in [-13,-4] \, a_0$.
    (e), (f): 
    Conditional wavefunctions $\phi(x_2;x_1)$ corresponding to the panels above, shown as contour plots (color indicates the sign).
    The states ``0'' and ``1'' discussed in Sec.\ \ref{sec:twostatemodel} in the context of the geometric potential $v^{\rm G}$ are marked.
    }
    \label{fig:pic_geo_asy}
  \end{figure}
  
  Fig.\ \ref{fig:pic_geo_asy}a shows the KS potential together with the relevant KS orbitals and Fig.\ \ref{fig:pic_geo_asy}b shows the EEF potential together with the one-electron density for an internuclear distance of \unit[$R = 5$]{$a_0$}.
  The lowest KS orbital $\vp_0^{\rm KS}$ is localized around \unit[$x_1 = -2.5$]{$a_0$} (where the nucleus with charge +2 is) while the highest occupied KS orbital $\vp_1^{\rm KS}$ is localized around \unit[$x_1 = +2.5$]{$a_0$} (where the nucleus with charge +1 is).
  
  The contributions to the EEF and KS potentials are depicted in panels Fig.\ \ref{fig:pic_geo_asy}c and \ref{fig:pic_geo_asy}d for the internuclear distances \unit[$R = 2$]{$a_0$} and \unit[$R = 5$]{$a_0$}, respectively.
  Also this model is well-described by Hartree-Fock theory, hence the relations \eqref{eq:pksimfs} hold.
  For larger distances, $v^{\rm H}$ and $v^{\rm HXC} + v^{\rm PH}$ form as step, because the state of the environment changes along $x_1$.
  For a detailed discussion of this step feature, see \cite{complet}.
  
  Here, we focus on the the geometric potential $v^{\rm G}$.
  It looks similar to that of the homonuclear diatomic, with a bell-shaped maximum centered at $x_1 \approx 0$ that indicates a qualitative change of the conditional wavefunction $\phi(x_2;x_1)$, depicted in Fig.\ \ref{fig:pic_geo_asy}e and \ref{fig:pic_geo_asy}f, respectively.
  However, for larger internuclear distances there is a second significant change of $\phi(x_2;x_1)$ along $x_1$, shown in the inset of \ref{fig:pic_geo_asy}d.
  For \unit[$R = 5$]{$a_0$} this second change is at \unit[$x_1 \approx -8$]{$a_0$}.
  If an electron is found somewhere in $-8 \, a_0 < x_1 < 0$, we see from $\phi(x_2;x_1)$ that the second electron is most probably found around \unit[$x_2 = 2.5$]{$a_0$}, corresponding to the location of the nucleus with charge $+1$.
  In contrast, if an electron is found at \unit[$x_1 < -8$]{$a_0$}, the second electron is found around \unit[$x_2 = -2.5$]{$a_0$}, corresponding to the nucleus with charge $+2$.
  The larger the internuclear distance $R$ becomes, the more does this transition move to smaller values of $x_1$ (i.e., to the left).
  Mechanistically, what happens at the location of the peak is a charge transfer, where the electron at $x_2$ switches from the energetically  higher potential well to the lower well depending on where the other electron at $x_1$ is located.  \cite{complet}
  Close inspection of the potentials for \unit[$R = 5$]{$a_0$} reveals that this charge transfer, which is a qualitative change in the conditional wavefunction $\phi(x_2;x_1)$, is visible in the $v^{\rm G}$, albeit barely: 
  The change happens over a rather large region along $x_1$ compared with the change at $x_1 \approx 0$ and thus leads only to a small but broad local increase in $v^{\rm G}$, sbown in the inset of \ref{fig:pic_geo_asy}d.
  \red{This second peak in $v^{\rm G}$ has also been found in a similar model recently \cite{giarrusso2018}, where it was concluded that the peak is at the side of the more electronegative atom, in agreement with our interpretation.
  We note that although there is a large difference in the height and with of the peak, the two changes of the conditional wavefunction are rather similar.}
  This leads to the integrals of $\sqrt{v^{\rm G}}$ over the corresponding regions to both be approximately equal, as explained in the next section.
  
  \begin{figure}[!htbp]
    \centering
    \includegraphics[width=.5\linewidth]{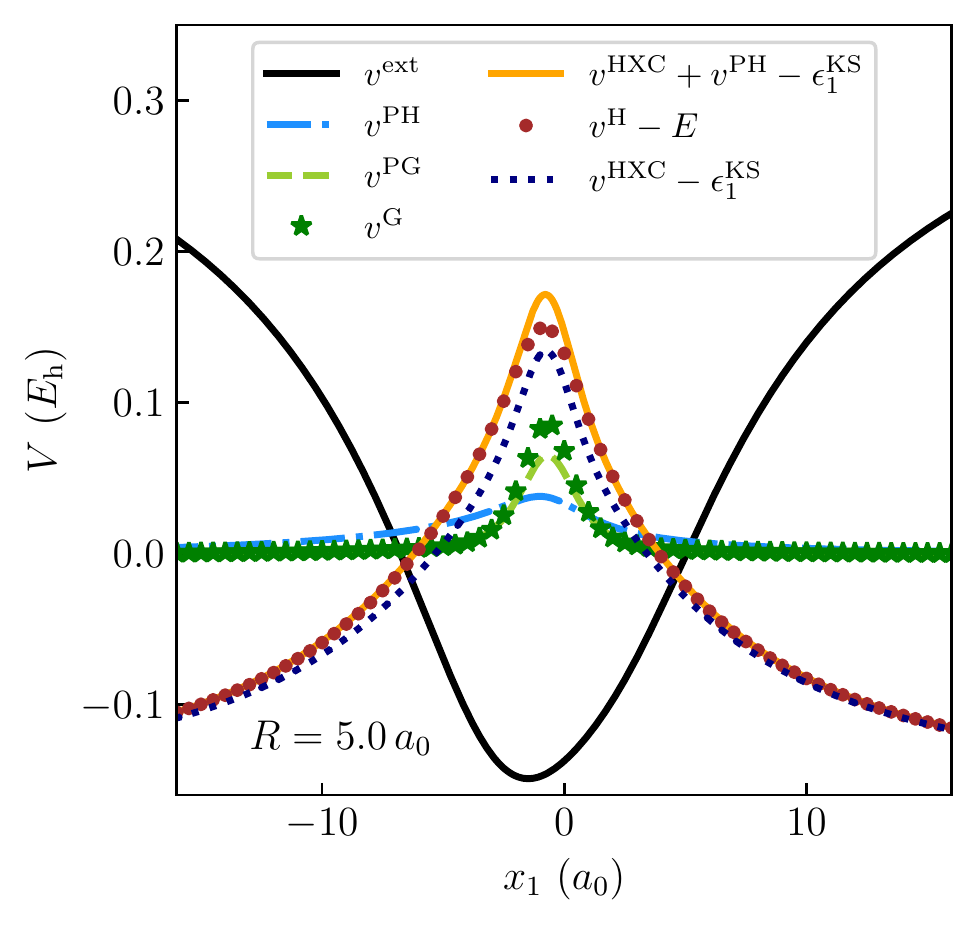}
    \caption{
    Like Fig.\ \ref{fig:pic_geo_asy}d, but for a very delocalized electron-nuclear interaction (parameter \unit[$c_{\rm en} = 25$]{$a_0^2$}).
    }
    \label{fig:pic_non_hf_only_pot}
  \end{figure}
  
  Before starting with a two-state analysis of the model, we look at what happens if there is a significant difference between the exact wavefunction and the Hartree-Fock wavefunction.
  In our experience, for one-dimensional models Hartree-Fock theory describes the electronic structure well and it its hard to make it ``fail''.
  \red{One possibility would be to consider the limit of strictly correlated electrons, as was done in \cite{giarrusso2018}.
  Here, we take a different approach and make} the electron correlation in our model hamiltonian more important by setting $c_{\rm en}$ to a large value, which results in a broad external potential $v_{\rm en}$ while keeping the electron-electron interaction $v_{\rm ee}$ sharply localized.
  In Fig.\ \ref{fig:pic_non_hf_only_pot}, we show the EEF and DFT potentials for such a large value.
  We find that the geometric potential $v^{\rm G}$ of the interacting wavefunction is higher that the geometric potential $v^{\rm PG}$ of the KS wavefunction, indicating that there is a stronger change of the conditional wavefunction along $x_1$ in the internuclear region for the interacting system.
  As $v$ is the same for the interacting and the KS system, this has to be compensated by $v^{\rm H}$ being smaller than $v^{\rm PH} + v^{\rm HXC}$, i.e., the energy of the environment is lower for the interacting than for the KS system.
  However, this behavior need \red{not always be} the case and we cannot make general statements about the energetic order of $v^{\rm G}$ compared to $v^{\rm PG}$ at the moment.
  
  \subsection{A two-state analysis}
  \label{sec:twostatemodel}
  
  The two parts $v^{\rm H}$ and $v^{\rm G}$ of the EEF potential $v$ are functionals of the conditional wavefunction $\phi$.
  By looking at the conditional wavefunctions $\phi$ for the symmetric and asymmetric diatomic in Fig.\ \ref{fig:pic_geo_sym} and in Fig.\ \ref{fig:pic_geo_asy}, respectively, is seems that $\phi(x_2;x_1)$ for some value of $x_1$ can be described as a two-state problem.
  This view is also supported by the observation that all our models are well-described with the Hartree-Fock approximation and the Hartree-Fock and Kohn-Sham orbitals differ very little.
  Hence, the wavefunction of the presented two-electron models of the diatomic molecule can approximately be written as
  \begin{align}
    \psi(x_1,x_2) \approx \frac{1}{\sqrt{2}} \left( \vp_0^{\rm KS}(x_1) \vp_1^{\rm KS}(x_2) 
                                                  - \vp_1^{\rm KS}(x_1) \vp_0^{\rm KS}(x_2) \right)
                                                  \label{eq:hfapprox}
  \end{align}
  with the conditional wavefunction
  \begin{align}
    \phi(x_2;x_1) \approx \frac{ \vp_0^{\rm KS}(x_1) \vp_1^{\rm KS}(x_2) 
                                                  - \vp_1^{\rm KS}(x_1) \vp_0^{\rm KS}(x_2) }{\sqrt{|\vp_0^{\rm KS}(x_1)|^2 + |\vp_1^{\rm KS}(x_1)|^2}}
                                                  \label{eq:hfapproxc}
  \end{align}
  for the gauge $\chi(x_1) = \sqrt{\rho(x_1)}$.
  Thus, for a given value of $x_1$, $\phi(x_2;x_1)$ is approximately a superposition of $\vp_0^{\rm KS}(x_2)$ and $\vp_1^{\rm KS}(x_2)$.
  
  This allows us to illustrate how the change of $\phi(x_2;x_1)$ along $x_1$ is encoded in the geometric potential $v^{\rm G}(x_1)$ in detail.
  The width and height of the bell-shaped maxima of $v^{\rm G}$ are closely related to the underlying structure of the conditional wavefunction $\phi(x_2;x_1)$ as a manifold with its own metric. 
  To express this metric explicitly for a general two-state system, we use that any linear combination of two orthonormal states $\ket{0}$ and $\ket{1}$ represents a state
  \begin{equation}
    \ket{\phi} (\theta,\varphi) = \cos(\theta/2) \ket{0} + e^{i \varphi} \sin(\theta/2) \ket{1} \, ,
  \end{equation}
  that is fully described with the two parameters $\theta \in [0,\pi]$ and $\varphi \in [0,2 \pi)$. 
  A geometrical representation of the state $\ket{\phi}$ is the Bloch sphere (Fig.\ \ref{fig:bloch01}) given by the two angles $\theta, \varphi$.
  The Fubini-Study metric \cite{provost1980} defines a distance between states.
  For a two-level system, expressed in terms of the parameters, it measures distances on the Block sphere and is given by
  \begin{equation}
    d s^2 = d \theta^2 + \sin^2 \theta \, d \varphi^2.
  \end{equation}
  
  \begin{figure}[!tbp]
    \centering
    \begin{minipage}[b]{0.49\textwidth}
      \centering
      \includegraphics[scale=1]{bloch-0.mps}
      \caption{Bloch sphere with coordinates $\theta$ and $\varphi$ spanning the parameter space of a superposition of two states.}
      \label{fig:bloch01}
    \end{minipage}
    \hfill
    \begin{minipage}[b]{0.49\textwidth}
      \centering
      \includegraphics[scale=1]{bloch-1.mps}
      \caption{Bloch sphere with coordinate $\vartheta$ spanning the parameter space of a superposition of two states if constrained to be real-valued.}
      \label{fig:bloch02}
    \end{minipage}
  \end{figure}
  
  If the conditional wavefunction $\phi(x_2; x_1)$ is a manifold of two states given by
  \begin{equation}
    \phi(x_2; x_1) = \braket{x_2|\phi} (\theta (x_1),\varphi (x_1)) \, ,
  \end{equation}
  the geometric potential $v^{\rm G}$ corresponds to the scaled square of the distance traversed on the Bloch sphere, which in terms of the $x_1$-dependent parameters $\theta$ and $\varphi$ is
  \begin{equation}
    v^{\rm G} (x_1) = \frac{1}{8} \left[ \left(\pa \theta \right)^2 + \sin^2 \theta \left(\pa \varphi \right)^2 \right] \, .
  \end{equation}
  Knowing that the conditional wavefunction $\phi(x_2; x_1)$ is real-valued reduces the configuration space of the whole Bloch sphere to states lying on the great circle.
  Thus, instead of $(\theta, \varphi)$ we use the central angle $\vartheta \in [0,2\pi)$ shown in Fig.\ \ref{fig:bloch02} to fully specify the conditional wavefunction for each value of $x_1$.

  For the states $\ket{0}$ and $\ket{1}$ we can choose the KS orbitals $\varphi^{\rm KS}_0$ and $\varphi^{\rm KS}_1$. 
  Then, the map between electron coordinate $x_1$ and the parameter $\vartheta$ is
  \begin{equation}
    \sin \vartheta (x_1) = \frac{- 2 \varphi^{\rm KS}_0(x_1) \varphi^{\rm KS}_1(x_1)}{\varphi^{\rm KS}_0(x_1)^2 + \varphi^{\rm KS}_1(x_1)^2} \, , ~~~~ \cos \vartheta (x_1) = \frac{\varphi^{\rm KS}_0(x_1)^2 - \varphi^{\rm KS}_1(x_1)^2}{\varphi^{\rm KS}_0(x_1)^2 + \varphi^{\rm KS}_1(x_1)^2} \, .
    \label{eq:thetav2}
  \end{equation}
  In the EEF picture the geometric potential $v^{\rm G}$ on the great circle is
  \begin{equation}
    v^{\rm G} (x_1) = \frac{1}{8} \left(\pa \vartheta\right)^2 \, ,
  \end{equation}
  which gives us another relation for the central angle,
  \begin{equation}
    \vartheta(x_1) = \int_{-\infty}^{x_1} \sqrt{8 v^{\rm{G}} (x_1')} \mathop{d x_1'} \, .
    \label{eq:theta}
  \end{equation}
  
  \begin{figure}[!htbp]
    \centering
    \includegraphics[width=.99\linewidth]{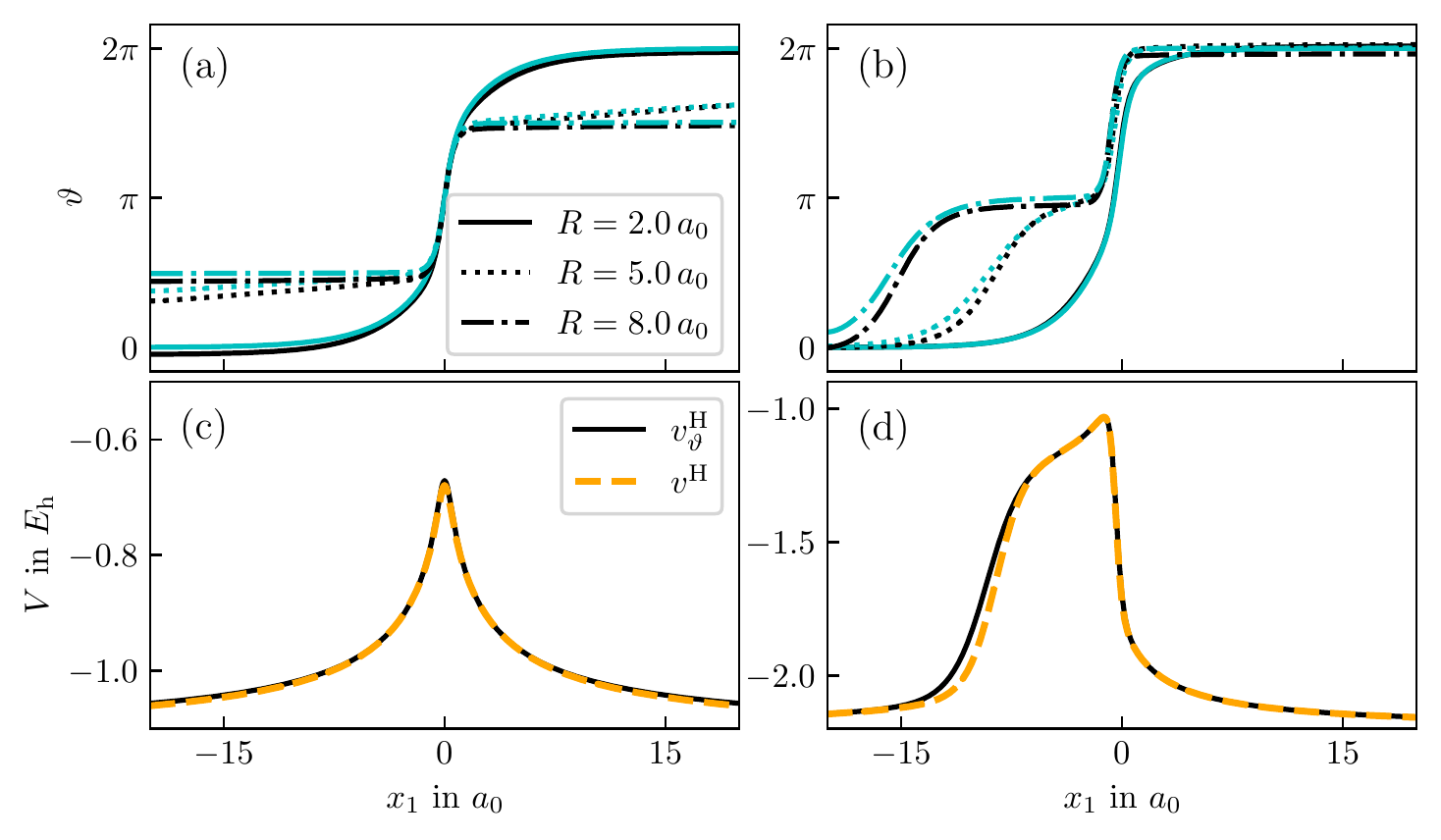}
    \caption{
    Top: Angle $\vartheta$ determined from \eqref{eq:theta} (black lines) and from \eqref{eq:thetav2} (cyan lines) for different values of the internuclear separation $R$ for (a) the homonuclear diatomic and (b) the heteronuclear diatomic.
    Bottom: Energy of the environmental electron $v^{\rm H}$ and $v_{\rm \vartheta}^{\rm H}$ determined by \eqref{eq:vHrec} with $\vartheta$ determined by \eqref{eq:thetav2}, for \unit[$R=5.0$]{$a_0$}, for (c) the homonuclear diatomic and (d) the heteronuclear diatomic.}
    \label{fig:pic_state_angle}
  \end{figure}
  
  There are thus two ways to compute the parameter $\vartheta(x_1)$ that determines the conditional wavefunction $\phi(x_2; x_1)$ in the two-state model:
  One is based on the KS orbitals, \eqref{eq:thetav2}, and one is based on the geometric potential, \eqref{eq:theta}.
  The parameter $\vartheta$ obtained in both ways is shown in Fig.\ \ref{fig:pic_state_angle}a and Fig.\ \ref{fig:pic_state_angle}b for the homonuclear and heteronuclear diatomic, respectively.
  There is little difference between the two ways of obtaining $\vartheta$ for both the model of the homo- and heteronuclear diatomic, i.e., the two-state approximation is justified.
  
  The behavior of $\vartheta$ reflects the behavior of $\phi$ depicted in Figs.\ \ref{fig:pic_geo_sym} and \ref{fig:pic_geo_asy}.
  To interpret $\phi$, some regions in these figures are marked with ``$0$'', ``$1$'', ``$+$'' or ``$-$''.
  These represent states between which $\phi(x_2; x_1)$ changes along $x_1$.
  For the homonuclear diatomic with small internuclear distance, $\phi(x_2; x_1)$ for $|x_1| \rightarrow \infty$ is close to the energetically lowest KS orbital $\vp_0^{\rm KS}$ and resembles the wavefunction of the ionized system.
  At $x_1 = 0$, it follows from \eqref{eq:hfapproxc} and $\vp_1^{\rm KS}(0) = 0$ that $\phi(x_2; x_1=0) \approx \vp_1^{\rm KS}(x_2)$.
  Hence, we have that $\phi(x_2; x_1)$ changes as 
  \begin{align}
    \phi(x_2; -\infty) \approx \vp_0^{\rm KS} 
      ~ \Rightarrow ~ \phi(x_2; 0) \approx \vp_1^{\rm KS}
      ~ \Rightarrow ~ \phi(x_2; +\infty) \approx \vp_0^{\rm KS}
      \label{eq:pathsym1}
  \end{align}
  (the sign of the state does not matter in the metric).
  This behavior is valid e.g.\ for an internuclear distance \unit[$R=2$]{$a_0$}, see Fig.\ \ref{fig:pic_geo_sym}e, where state ``0'' is approximately $\vp_0^{\rm KS}$ and ``1'' is approximately $\vp_1^{\rm KS}$.
  The path \eqref{eq:pathsym1} of $\phi$ from $x_1 = -\infty$ to $x_1 = +\infty$ corresponds to one full motion along the great circle, i.e., to a total change 
  \begin{align}
    |\vartheta(x_1=+\infty) - \vartheta(x_1=-\infty)| \approx 2 \pi
  \end{align}
  of the parameter $\vartheta$, as shown in Fig.\ \ref{fig:pic_state_angle}a for \unit[$R=2$]{$a_0$}.
  
  In contrast, for larger internuclear distances, $\phi(x_2; x_1)$ for $|x_1| \rightarrow \infty$ is either located on one or the other nucleus.
  Due to the symmetry of the problem, this corresponds approximately to the two states 
  \begin{align}
   \vp_{\pm}^{\rm KS} = \frac{1}{\sqrt{2}} \left( \vp_0^{\rm KS} \pm \vp_1^{\rm KS} \right)
  \end{align}
  which are on opposite sides of the Bloch sphere at the equator ($\ket{\pm}$ in Fig.\ \ref{fig:bloch02}), if $\vp_0^{\rm KS}$ and $\vp_1^{\rm KS}$ are the poles ($\ket{0}$ and $\ket{1}$). 
  Equation \eqref{eq:hfapproxc} is still valid and $\phi(x_2; x_1=0) \approx \vp_1^{\rm KS}(x_2)$, hence we have 
  \begin{align}
    \phi(x_2; x_1 \ll 0) \approx \vp_{+}^{\rm KS}
      ~ \Rightarrow ~ \phi(x_2; 0) \approx \vp_1^{\rm KS}
      ~ \Rightarrow ~ \phi(x_2; x_1 \gg 0) \approx \vp_{-}^{\rm KS}
      \label{eq:pathsym2}
  \end{align}
  which is visible in Fig.\ \ref{fig:pic_geo_sym}f, where state ``$+$'' is approximately $\vp_{+}^{\rm KS}$ and state ``$-$'' is approximately $\vp_{-}^{\rm KS}$.
  The path \eqref{eq:pathsym2} corresponds to half of a great circle on the Bloch sphere, i.e.,
  \begin{align}
    |\vartheta(x_1 \gg 0) - \vartheta(x_1 \ll 0)| \approx \pi,
  \end{align}
  as shown in Fig.\ \ref{fig:pic_state_angle}a for \unit[$R=5$]{$a_0$} and \unit[$8$]{$a_0$}.
  
  For the heteronuclear diatomic at small internuclear distances the situation is similar to that of the homonuclear diatomic, see Fig.\ \ref{fig:pic_geo_asy}e.
  In contrast, at larger internuclear distances, the KS orbitals are either localized on one or the other nucleus.
  From Fig.\ \ref{fig:pic_geo_asy}f we can see that there is a hopping of $\phi(x_2;x_1)$ from one side to the other and back along $x_1$, as discussed above, which is
  \begin{align}
    \phi(x_2; -\infty) \approx \vp_0^{\rm KS}
      ~ \Rightarrow ~ \phi(x_2; x_{\rm S}) \approx \vp_1^{\rm KS}
      ~ \Rightarrow ~ \phi(x_2; +\infty) \approx \vp_0^{\rm KS}
  \end{align}
  with $x_{\rm S}$ representing the corresponding region of $x_1$.
  Thus, 
  \begin{align}
    |\vartheta(x_1=+\infty) - \vartheta(x_1=-\infty)| \approx 2 \pi.
  \end{align}
  for any finite value of the internuclear distance $R$ but happens in two steps which each correspond to a change of $\pi$, with one smooth change somewhere at $x_1<0$ and a sharp change at $x_1\approx0$, as can be seen in Fig.\ \ref{fig:pic_state_angle}b.
  The smooth change moves more and more towards negative $x_1$ when the internuclear distance is increased.
  
  The fact that the asymptotic behavior of the KS orbitals is known can be used to construct $v^{\rm G}$ analytically in the region where the dominant orbital changes, at least if this asymptotic behavior is attained.
  The leading term in the asymptotic KS orbitals is
  \begin{equation}
    \varphi^\mathrm{KS}_i (x_1) \propto e^{-\sqrt{-2 \varepsilon^\mathrm{KS}_i} |x_1|}.
    \label{eq:ksorbasy}
  \end{equation}
  The dominant orbital changes, say, around $x_{\rm c}$, and we write the KS orbitals as
  \begin{equation}
    \varphi_i (x_1) = C e^{-A_i (x_1-x_{\rm c})} \, ,
  \end{equation}
  where $A_i$ is $\pm \sqrt{-2 \varepsilon^\mathrm{KS}_i}$ with the sign depending on the direction of the decay. 
  Using the explicit form of the KS orbitals in \eqref{eq:thetav2} leads to
  \begin{equation}
    \cos \vartheta = \frac{1 - e^{-2 \Delta (x_1-x_{\rm c})}}{1 + e^{-2 \Delta (x_1-x_{\rm c})}} \, , ~~~~~~~~ \sin \vartheta = \frac{2 e^{-\Delta (x_1-x_{\rm c})}}{1 + e^{-2 \Delta (x_1-x_{\rm c})}} \, ,
  \end{equation}
  where $\Delta = A_i - A_j$ with $i$ and $j$ being the indices of the two involved KS orbitals.
  We thus find that the geometric potential is a bell-shaped function with the width $1/\Delta$ and the height $\Delta^2/8$,
  \begin{equation}
    v^{\rm G} (x_1) = \frac{1}{8} \left(\pa \vartheta \right)^2 =   \frac{1}{8} \left( \frac{1}{ \sin \vartheta} \pa \cos (\vartheta) \right)^2 = \frac{1}{8} \Delta^2 \sech^2(\Delta (x_1-x_{\rm c})) \, .
    \label{eq:vga}
  \end{equation}
  The integral
  \begin{equation}
    \int \sqrt{8 v^{\rm G} (x_1)} \, dx_1 = \gd (\Delta (x_1-x_{\rm c})) \, ,
  \end{equation}
  is the Gudermannian function $\gd$ and yields
  \begin{equation}
    \int_{-\infty}^{+\infty} \sqrt{8 v^{\rm G} (x_1)} \, dx_1 = \pi.
    \label{eq:pipi}
  \end{equation}
   Relation \eqref{eq:pipi} means that the integral of $\sqrt{8 v^{\rm G}}$ over the region of a change of the KS orbital which dominates the density is equal to $\pi$.
   The analytic form \eqref{eq:vga} of $v^{\rm G}$ is not restricted to our models but is general, provided the KS orbitals can be described by \eqref{eq:ksorbasy} and provided there is only a switch from one dominant orbital to another, which is typically the case.
  
  If the internuclear distance is large enough, e.g.\ for \unit[$R=5$]{$a_0$} and for \unit[$R=8$]{$a_0$}, there are one or two such regions for the model of the homo- or heteronuclear diatomic molecule, respectively.
  For the heteronuclear diatomic, the shape of the geometric potential of the region outside the internuclear region is a broad and shallow bell with parameter $\Delta = \sqrt{-2 \varepsilon^\mathrm{KS}_0} - \sqrt{-2 \varepsilon^\mathrm{KS}_1}$, see the inset in Fig.\ \ref{fig:pic_geo_asy}d.
  For both the homo- and heteronuclear diatomic, the geometric potential in the internuclear region is a narrow and high bell, see Fig.\ \ref{fig:pic_geo_asy}c and \ref{fig:pic_geo_asy}d, respectively.
  If the internuclear distance $R$ is large enough, the corresponding parameter is $\Delta = \sqrt{-2 \varepsilon^\mathrm{KS}_0} + \sqrt{-2 \varepsilon^\mathrm{KS}_1}$ (for the homonuclear case also $\varepsilon^\mathrm{KS}_0 \approx \varepsilon^\mathrm{KS}_1$).
  The considered values of $R$ (up to ca.\ \unit[$R=10$]{$a_0$}, then the electron density in the internuclear region becomes too small and numerical artefacts appear) are still too small for this relation to hold, but \eqref{eq:vga} is valid, albeit with different $\Delta$.
  
  Finally, we note that we can use the two-state assumption to construct $v^{\rm H}$ from $v^{\rm G}$ (and vice versa).
  For this purpose, the parameter $\vartheta$ can be determined via \eqref{eq:theta} from the geometric potential.
  The energy of the environment is then given as
  \begin{equation}
  v^{\rm H} (x_1) \approx v_{\vartheta}^{\rm H} (x_1) 
  = \frac{h_{00} + h_{11}}{2} + \frac{h_{00} - h_{11}}{2} \cos \vartheta (x_1) +  h_{01} \sin \vartheta (x_1).
  \label{eq:vHrec}
  \end{equation}
  Here, $h_{ij}$ correspond to matrix elements of the operator evaluated for $v^{\rm H}$ in the chosen basis $\ket{0}$ and $\ket{1}$.
  In the basis of the KS orbitals and for the model potential, the matrix elements are
  \begin{align}
    h_{ij} = \Braket{\vp_i^{\rm KS}(2) | -\frac{\pb^2}{2} + v_{\rm en}(2) + v_{\rm ee}(1,2) | \vp_j^{\rm KS}(2)}_2.
  \end{align}
  The only $x_1$-dependence of $h_{ij}$ is due to the electron-electron interaction $v_{\rm ee}$.
  Fig.\ \ref{fig:pic_state_angle}c and \ref{fig:pic_state_angle}d illustrate that the reconstruction works excellent for \unit[$R=5$]{$a_0$}:
  From $v^{\rm G}$ the angle $\vartheta$ can be determined via \eqref{eq:theta}, and this angle can be used in \eqref{eq:vHrec} to obtain $v_{\vartheta}^{\rm H}$, which is very close to the true energy $v^{\rm H}$ of the environment.
  This reconstruction works equally well for the other considered internuclear distances.
  
  \section{Summary}
  
  The EEF has a clear picture as the description of an electron in the environment of other electrons.
  The wavefunction of those other electrons provides the scalar potential $v$ and vector potential that appear in the one-electron Schr\"odinger equation, where $v$ is the sum of the energy of the environment $v^{\rm H}$ in the presence of an additional electron and a geometric potential $v^{\rm G}$.
  In the EEF, the interacting many-electron wavefunction is considered.
  If the KS wavefunction is used instead, the same one-electron potential $v$ is obtained but the contributions to $v$ are different due to the different way of how the electronic structure is described.
  
  The connection between the EEF and DFT allows to interpret the KS potentials, i.e., the Hartree-exchange-correlation potential $v^{\rm HXC}$ and the Pauli potential $v^{\rm P}$, from the EEF perspective.
  In particular, we showed that $v^{\rm P}$ contains the geometric potential of the KS system and the energy of the KS environment, whereas $v^{\rm HXC}$ can be viewed as a correction to the external potential due to the different electron-electron interaction in the KS system compared to the interacting system.
  This has to be contrasted with the usual view of the Pauli potential as being the difference of a non-interacting fermionic and bosonic system -- from the EEF perspective, both $v^{\rm HXC}$ the Pauli potential describe the fermionic problem itself, just in a different way than how the interacting many-electron wavefunction describes the problem.
  
  In contrast to $v^{\rm H}$, the physical meaning of the geometric potential $v^{\rm G}$ is less obvious.
  We explained its connection to the quantum geometric tensor and studied its behavior for a model of a two-electron diatomic molecule.
  This model can be understood in a two-state picture, which allowed us to illustrate that $v^{\rm G}$ is proportional to the change of state of the environment depending on where the additional electron is.
  Also, we provided an analytical form as well as constraints on the integral of $\sqrt{v^{\rm G}}$ if a state change happens.
  These results may be useful also for more general systems if the state change is that of one KS orbital to another.
  Such situations are essentially charge transfers and they are common in molecules, hence further investigation on the behavior of $v^{\rm G}$ can help to model its contribution in the EEF or in OF-DFT.
  
  What we largely ignored in our discussion is the role of the vector potential.
  Although it may not be needed to describe the ground state of a many-electron system, it will certainly be relevant for rotating molecules, molecules in laser fields and possibly for describing degenerate states in molecules.
  To investigate the relevant of the vector potential it is, however, necessary to look at three-dimensional model systems. 
  Suitable systems are much harder to find and to simulate, but it poses an interesting challenge for future work.
  
  \section*{Acknowledgement}
    
    AS thanks Denis Jelovina (ETH Z\"urich) for helpful discussions.
    This research is supported by an Ambizione grant of the Swiss National Science Foundation (SNF).
  
  \bibliography{lit}{}
  \bibliographystyle{unsrt}

  \appendix
  
    \section{Other expressions for the potentials in the EEF}
      \label{sec:eefpot}
    
      Relations \eqref{eq:eef_vt}, \eqref{eq:eef_vv}, and \eqref{eq:eef_vfs} for the potentials $v^{\rm T}$, $v^{\rm V}$, and $v^{\rm G}$, respectively, can be expressed in terms of the many-electron wavefunction $\psi$ and the one-electron density $\rho = |\chi|^2$ by using the relation \eqref{eq:phi} for the conditional wavefunction $\phi$.
      We first note that the geometric potential is also given by 
      \begin{align}
        v^{\rm G}(\vec{r}) 
          &= \frac{1}{2} \braket{\prv \phi(2,\dots,N;\vec{r}) | \hat{P} | \prv \phi(2,\dots,N;\vec{r})}_{2 \dots N},
          \label{eq:eef_vfs2}
      \end{align}
      where 
      \begin{align}
        \hat{P} = 1 - \ket{\phi(2,\dots,N;\vec{r})}\bra{\phi(2,\dots,N;\vec{r})}
      \end{align}
      is a projection operator on the state orthogonal to $\phi$ for a given value of $\vec{r}$.
      Expression \eqref{eq:eef_vfs2} shows the close connection to the Fubini-Study metric \cite{provost1980} and the geometric meaning of $v^{\rm G}$.
      
      Using \eqref{eq:phi}, it is straightforward to show that 
      \begin{align}
        v^{\rm T}(\vec{r})  
          &= \frac{1}{\rho(\vec{r})} 
            \matrixel{\psi(2,\dots,N;\vec{r})}{-\sum\limits_{j=2}^N \frac{\nabla_j^2}{2}}{\psi(2,\dots,N;\vec{r})}_{2 \dots N} \label{eq:vt_psi} \\
        v^{\rm V}(\vec{r})  
          &= \frac{1}{\rho(\vec{r})}
            \matrixel{\psi(2,\dots,N;\vec{r})}{V(1,2,\dots,N)}{\psi(2,\dots,N;\vec{r})}_{2 \dots N} - v^{\rm ext}(\vec{r})
            \label{eq:vv_psi} \\
        v^{\rm G}(\vec{r}) 
          &= \frac{1}{2 \rho(\vec{r})} \braket{\prv \psi(2,\dots,N;\vec{r}) | \hat{P}_{\psi} | \prv \psi(2,\dots,N;\vec{r})}_{2 \dots N},
          \label{eq:vg_psi}
      \end{align}
      with 
      \begin{align}
        \hat{P}_{\psi} = 1 - \frac{1}{\rho(\vec{r})} \ket{\psi(2,\dots,N;\vec{r})}\bra{\psi(2,\dots,N;\vec{r})}.
      \end{align}
      Also $\hat{P}_{\psi}$ is a projector for a given value of $\vec{r}$, i.e., in the subspace of the coordinates $\vec{r}_2, \dots, \vec{r}_N$.
      As only $\psi$ and $\rho$ appear in \eqref{eq:vt_psi}, \eqref{eq:vv_psi}, and \eqref{eq:vg_psi}, but not $\chi$, it is clear that $v^{\rm T}$, $v^{\rm V}$, and $v^{\rm G}$ do not depend on the gauge, i.e., on the choice of the phase of $\chi$.
    
%     \section{Additional information for the atomic model}
%       \label{sec:addinfo}
%     
%       In the discussion of the one-dimensional model of an atom in section \ref{sec:models}, it was mentioned that we also studied two states of the three-electron Hamiltonian $H_{\rm a}^{(3)}$.
%       Those states are the energetically lowest doublet state and the lowest fully antisymmetric eigenstate.
%       
%       \begin{figure}[!htbp]
%         \centering
%         \includegraphics[width=.8\linewidth]{pic_3e1n1d_t_pot1.pdf}
%         \caption{Contributions to the potential $v$ in the EEF and in OF-DFT for the energetically lowest antisymmetric eigenstate of the model of a three-electron atom, for two values of the electron-electron interaction parameter $\lambda$.}
%         \label{fig:pic_3e1n1d_t_pot1}
%       \end{figure}
%       
%       The contributions to $v$ for theses states are very similar to the contributions for the lowest antisymmetric eigenstate of the two-electron.
%       Fig.\ \ref{fig:pic_3e1n1d_t_pot1} illustrates this for the lowest fully antisymmetric eigenstate of $H_{\rm a}^{(3)}$, for two different values of the parameter $\lambda$ which controls the electron-electron interaction.
    
    \section{The geometric potential and the kinetic energy density}
      \label{sec:ekind}
      
      The difference between the geometric potential $v^{\rm G}$ of the interacting system and the geometric potential $v^{\rm PG}$ of the KS system can, by construction, be related to the different kinetic energy densities.
      For the gauge $\vec{A} \stackrel{!}{=} 0$ and for the case $\chi = \sqrt{\rho} \in \mathbb{R}$, $\phi \in \mathbb{R}$, we have 
      \begin{align}
        v^{\rm G}(\vec{r})
          &= \frac{1}{2} \braket{\left(\prv \phi(2,\dots,N|\vec{r})\right)^2}_{2 \dots N}
          = \frac{1}{2} \Braket{\left(\prv \frac{\psi(\vec{r},2,\dots,N)}{\chi(\vec{r})}\right)^2}_{2 \dots N} \\
          &= \frac{1}{2} 
              \left(  \frac{\braket{(\prv \psi)^2}_{2 \dots N}}{\chi^2} 
                      - \frac{\braket{\prv (\psi^2)}_{2 \dots N}}{\chi^3} \prv \chi
                      + \frac{(\prv \chi)^2}{\chi^2}
              \right) \\
          &=  \frac{t(\vec{r})}{\chi^2} 
          - \frac{1}{2} \frac{\prv \rho}{\rho} \frac{\prv \chi}{\chi} 
          + \frac{1}{2} \frac{(\prv \chi)^2}{\chi^2},
          \label{eq:vge}
      \end{align}
      where 
      \begin{align}
        t(\vec{r}) = \frac{\braket{(\prv \psi)^2}_{2 \dots N}}{2} 
      \end{align}
      is the positive-defined one-electron kinetic energy density of the interacting system.
      For the KS system a similar relation holds, i.e., 
      \begin{align}
        v^{\rm PG}(\vec{r}) 
          &= \frac{1}{2} \sum_{n=1}^N \left( \prv \frac{\vp_n^{\rm KS}(\vec{r})}{\chi(\vec{r})} \right)^2
          = \frac{1}{2} \sum_{n=1}^N 
              \left( \frac{\prv \vp_n^{\rm KS}}{\chi} - \frac{\vp_n^{\rm KS} \prv \chi}{\chi^2}
              \right)^2 \\
          &= \frac{1}{2} \sum_{n=1}^N  \left( \frac{(\prv \vp_n^{\rm KS})^2}{\chi^2} - \frac{\prv \left((\vp_n^{\rm KS})^2\right)}{\chi^3} \prv \chi + \frac{(\vp_n^{\rm KS})^2 }{\chi^4} (\prv \chi)^2 \right) \\
          &= \frac{t^{\rm KS}(\vec{r})}{\chi^2} - \frac{1}{2} \frac{\prv \rho}{\rho} \frac{\prv \chi}{\chi} + \frac{1}{2} \frac{(\prv \chi)^2}{\chi^2} 
          \label{eq:vgpe}
      \end{align}
      where 
      \begin{align}
        t^{\rm KS}(\vec{r}) = \frac{\sum_{n=1}^N (\prv \vp_n^{\rm KS})^2}{2}
      \end{align}
      is the positive-defined one-electron kinetic energy density of the KS system.
      Clearly, \eqref{eq:vge} and \eqref{eq:vgpe} differ only by two kinetic energy densities $t$ and $t^{\rm KS}$.
      
\end{document}